\PassOptionsToPackage{unicode}{hyperref}
\PassOptionsToPackage{hyphens}{url}
\documentclass[
]{article}
\usepackage{amsmath,amssymb}
\usepackage{cmap}          % searchable/copyable PDF text
\usepackage{mathptmx}      % Times text + maths (arXiv-safe Type1)
\usepackage{iftex}
\ifPDFTeX
  \usepackage[T1]{fontenc}
  \usepackage[utf8]{inputenc}
  \usepackage{textcomp} % provide euro and other symbols
\else % if luatex or xetex
  \usepackage{unicode-math} % this also loads fontspec
  \defaultfontfeatures{Scale=MatchLowercase}
  \defaultfontfeatures[\rmfamily]{Ligatures=TeX,Scale=1}
\fi
\IfFileExists{lmodern.sty}{\usepackage{lmodern}}{\PassOptionsToPackage{expansion=false}{microtype}}
\ifPDFTeX\else
\fi
\IfFileExists{upquote.sty}{\usepackage{upquote}}{}
\IfFileExists{microtype.sty}{% use microtype if available
  \usepackage[]{microtype}
  \UseMicrotypeSet[protrusion]{basicmath} % disable protrusion for tt fonts
}{}
\makeatletter
\@ifundefined{KOMAClassName}{% if non-KOMA class
  \IfFileExists{parskip.sty}{%
    \usepackage{parskip}
  }{% else
    \setlength{\parindent}{0pt}
    \setlength{\parskip}{6pt plus 2pt minus 1pt}}
}{% if KOMA class
  \KOMAoptions{parskip=half}}
\makeatother
\usepackage{xcolor}
\usepackage{longtable,booktabs,array}
\usepackage{calc} % for calculating minipage widths
\usepackage{etoolbox}
\makeatletter
\patchcmd\longtable{\par}{\if@noskipsec\mbox{}\fi\par}{}{}
\makeatother
\IfFileExists{footnotehyper.sty}{\usepackage{footnotehyper}}{\usepackage{footnote}}
\makesavenoteenv{longtable}
\usepackage{graphicx}
\makeatletter
\def\maxwidth{\ifdim\Gin@nat@width>\linewidth\linewidth\else\Gin@nat@width\fi}
\def\maxheight{\ifdim\Gin@nat@height>\textheight\textheight\else\Gin@nat@height\fi}
\makeatother
\setkeys{Gin}{width=\maxwidth,height=\maxheight,keepaspectratio}
\makeatletter
\def\fps@figure{htbp}
\makeatother
\ifLuaTeX
  \usepackage{selnolig}  % disable illegal ligatures
\fi
\IfFileExists{bookmark.sty}{\usepackage{bookmark}}{\usepackage{hyperref}}
\IfFileExists{xurl.sty}{\usepackage{xurl}}{} % add URL line breaks if available
\hypersetup{
  pdftitle={Settlement Infrastructure, Inside Money Elasticity, and the Network Economics of Distributed Ledger Technology},
  hidelinks,
  pdfcreator={LaTeX via pandoc}}

\title{Settlement Infrastructure, Inside Money Elasticity, and the Network Economics of Distributed Ledger Technology}
\author{Michail Samawi \quad Hui Gong \quad Francesca Medda \\ \smallskip \small University College London -- Institute of Finance and Technology}
\date{}

\usepackage{amsmath,amssymb}
\usepackage{booktabs}
\usepackage{longtable}
\usepackage{array}
\usepackage{caption}
\usepackage{rotating}
\usepackage{microtype}
\usepackage[margin=1in]{geometry}
\usepackage{setspace}
\usepackage{etoolbox}
\usepackage{placeins}
\usepackage{float}
\AtBeginEnvironment{longtable}{\footnotesize}
\begin{document}
\maketitle
\begin{abstract}
{We construct the Settlement Modernisation Index, a panel dataset of 809 reform events across 24 advanced economies between 1993 and 2024, decomposed into three economic channels and three adoption phases. We document an S-curve in inside money elasticity with two interior turning points at SMI = 0.27 and 0.93, separating a liberation phase, a post-global-financial-crisis compliance valley, and a mature-infrastructure recovery phase. We show that settlement modernisation generates network-conditional balance sheet efficiencies through a T2S event-study with year-by-year EMIR decomposition (saturation $\beta$ = +0.557, p \textless{} 0.01) and an out-of-sample synthetic control null on Switzerland's post-2021 SDX deployment. Applied along the BIS three-layer connectivity taxonomy, the framework forecasts +13.4 percent efficiency recovery from the ECB's Pontes initiative over 2027--2032. Conditional UK and US accession to the Appia composability layer (2028) raises the ceiling to +37.5 percent. Balance-sheet efficiencies from atomic settlement are a property of the bilateral pair, not the node.}


\end{abstract}

\section{I. Introduction}
The European Central Bank launches Project Pontes in the third quarter of 2026 to settle distributed-ledger transactions in central bank money. Complementing this initiative, Project Appia represents the longer-term strategic vision of developing an integrated tokenised financial ecosystem, capable of supporting seamless interaction between tokenised deposits, digital securities, and central bank settlement assets (Cipollone, 2026; ECB, 2025). By connecting tokenised deposits, digital securities, and central bank settlement mechanisms within a common infrastructure, these initiatives seek to enhance financial efficiency, strengthen monetary sovereignty, and support the development of a competitive European digital financial architecture.

The Bank of England launches its Synchronisation Lab the same year to integrate the RT2 real-time gross settlement system with external DLT ledgers, and plans the issuance of a digital gilt (DIGIT) on the same architecture (Breeden, 2025; Cleland, 2025). The Bank for International Settlements, through Project Agorá and its Innovation Hub, treats atomic settlement on unified ledgers as the analytical baseline for the next generation of wholesale payments (Auer et al., 2025; Carstens, 2023). The efficiency case across all three rests on a single premise: that atomic settlement eliminates the bilateral exposure window that has constrained wholesale cross-border banking since the 1974 Herstatt collapse (BIS, 1996)

Switzerland already ran the experiment. Its SIX Digital Exchange completed the world\textquotesingle s first operationally live atomic delivery-versus-payment settlement in 2022 --- the frontier of settlement modernisation, simultaneously addressing principal risk, liquidity prefunding, and collateral immobilisation in a single indivisible on-chain transaction. The measured effect on Switzerland\textquotesingle s inside money elasticity in 2022--2024 was statistically indistinguishable from zero. The most technologically advanced settlement economy in the panel produced no efficiency effect from its most advanced reform.

Nonetheless, the empirical measurement of cross-border wholesale settlement quality poses a fundamental challenge in the context of these emerging initiatives. Wholesale settlement infrastructures constitute institutional arrangements that evolve gradually and often through discrete regulatory and technological reforms rather than continuous adjustments. Consequently, assessing whether Projects Pontes and Appia will deliver the efficiency gains and welfare improvements anticipated by their proponents depends on institutional conditions that are not adequately captured by existing measures of settlement infrastructure.

This challenge is further compounded by the fact that settlement reforms occur infrequently and exhibit substantial heterogeneity across jurisdictions, rendering conventional time-series approaches of limited reliability. The existing literature has sought to address these limitations through event-study methodologies (Galati, 2002), cross-sectional comparisons of payment and settlement systems (BIS, 2012), and analyses of regulatory reforms and market structure (Pirrong, 2011). While these approaches provide valuable insights, they do not offer a comprehensive framework capable of jointly capturing the regulatory, technological, and market-practice dimensions that determine the effectiveness of wholesale settlement infrastructures over time.

The objective of this paper is to address this gap through a three-stage analytical framework. First, we construct the Settlement Modernisation Index (SMI), a novel longitudinal measure designed to capture the evolution of wholesale cross-border settlement infrastructures and their associated balance-sheet efficiencies across advanced economies since 1993. The index reveals a non-linear trajectory of settlement modernisation characterised by an S-shaped pattern comprising three distinct phases: an initial liberation phase associated with market integration and efficiency gains; a post-Global Financial Crisis (GFC) compliance phase marked by increased collateralisation and regulatory frictions; and a subsequent recovery phase driven by the maturation of market infrastructures and liquidity-saving technologies.

Second, we validate the underlying transmission mechanism through historical evidence drawn from the adoption of Delivery-versus-Payment (DvP) arrangements, Continuous Linked Settlement (CLS), and TARGET2-Securities (T2S). The empirical evidence indicates that settlement modernisation exhibits a network structure in which efficiency gains do not arise solely from technological implementation but from the achievement of critical mass across bilateral counterparty relationships. In other words, the welfare benefits of settlement reform materialise only when adoption reaches a sufficiently broad level of systemic market practice, generating network externalities that enhance liquidity, reduce settlement risk, and improve balance-sheet efficiency.

Third, building on this validated historical framework, we map the forthcoming transition toward distributed ledger technology (DLT)-based settlement infrastructures onto the three connectivity phases identified by the Bank for International Settlements (BIS). This approach allows us to trace the potential pathway from fragmented experimentation to fully integrated digital settlement ecosystems and to quantify the associated efficiency gains.

The first phase, Fragmentation, is characterised by isolated DLT platforms operating without meaningful interoperability. Consistent with this interpretation, evidence from the Switzerland SDX implementation, evaluated through a synthetic control framework, suggests negligible efficiency gains at this stage. The second phase, Interoperability, is expected to emerge with the implementation of Project Pontes, the European Central Bank\textquotesingle s initiative to connect DLT-based transactions to central bank settlement infrastructure through Hash Time-Locked Contract (HTLC) mechanisms linked to TARGET services. Calibrating this transition against the coordinated migration experience of T2S, our estimates suggest that interoperability could generate an efficiency recovery of approximately 13.4 per cent over the period 2027--2032.

The final phase, Composability, represents the emergence of an integrated tokenised financial ecosystem in which assets, deposits, collateral, and settlement functions interact seamlessly across jurisdictions and platforms. We illustrate this stage through the prospective architecture envisaged under Project Appia. Simulated accession scenarios indicate that the efficiency gains from composability are highly sensitive to the breadth of international participation. Conditional participation by the United Kingdom increases the projected efficiency recovery to 28.4 per cent, while the inclusion of both the United Kingdom and the United States raises the potential recovery to 37.5 per cent. These findings highlight the central role of cross-jurisdictional interoperability and composability in unlocking the full welfare benefits of next-generation settlement infrastructures and suggest that the economic value of DLT-based financial architectures derives not merely from technological innovation, but from the scale and integration of the networks they create.

The paper therefore provides contributions to three strands of literature. The monetary economics literature on the relationship between financial infrastructure and money demand (Gorton \& Pennacchi, 1990; Kahn \& Roberds, 2009) provides the theoretical foundation for treating settlement quality as a determinant of inside money productivity. The payment systems literature on settlement risk and its resolution (BIS, 1996; Duffie \& Zhu, 2011; Galati, 2002) establishes the channels through which infrastructure affects balance sheet intensity. The emerging literature on DLT in wholesale financial markets (BIS, 2021, 2023; ECB, 2024) motivates the forward-looking efficiency prediction. Our contribution to each strand is empirical: the SMI provides the first systematic measurement of settlement infrastructure development at the panel-country level, enabling estimation of efficiency effects that were previously documented only in case studies or theoretical frameworks.

The paper proceeds as follows. Section II develops the theoretical framework, deriving the three-channel balance sheet decomposition, the atomic settlement proof, and the Network Efficiency Variable formula. Section III describes the SMI construction and the phase-timing validation test that identifies systemic market practice, not regulatory enactment or technology deployment, as the moment efficiency materialises. Section IV presents the primary results: the S-curve, the two turning points, the three-channel decomposition, and the mandatory clearing divide that separates the compliance valley from the mature-infrastructure recovery phase. Section V presents four country portraits --- Luxembourg, Japan, Switzerland, and Korea --- that trace the channels through specific national histories. Section VI addresses causal identification through a sign-structure argument and a seven-specification robustness battery. Section VII presents the network efficiency prediction along the BIS three-layer connectivity taxonomy: Switzerland (Fragmentation), Pontes Q3 2026 (Interoperability), and Appia 2028 (Composability). Section VIII concludes.

\section{II. Theoretical Framework}

This section develops the theoretical framework in four steps. Section II.A establishes the balance-sheet constraint that settlement imposes, taking the Herstatt failure as the canonical case: the interval between payment and receipt is a window of full principal exposure, and the reserve held against it is a claim on capacity that could otherwise fund cross-border lending. Section II.B decomposes settlement reform into three institutional channels --- principal-at-risk, liquidity and payment finality, and collateral immobilisation --- and shows that they act on that constraint with opposing signs. Section II.C proves that atomic settlement eliminates the exposure window for a bilateral pair. Section II.D derives the Network Efficiency Variable, the measure of value that pairwise atomicity generates across a system.

\subsection{II.A The Herstatt Problem as a Balance Sheet Constraint}
On 26 June 1974, the Bundesbank withdrew the banking licence of Herstatt Bank at 3:30 in the afternoon, Frankfurt time. By that hour, several counterparty banks had already paid Deutsche Mark legs of foreign exchange transactions to Herstatt, expecting to receive the dollar counterparts in New York later the same day. When the doors closed, the dollar payments stopped. The banks that had already paid received nothing. The full principal value of their transactions --- not the mark-to-market exposure, not the net position, but the entire face value of every unsettled trade --- was gone (BIS, 1996).

The Herstatt collapse was not a large bank failure by the standards of financial history. Its direct losses were manageable. Its lasting consequence was a change in how every bank that transacted cross-border priced the interval between trade agreement and payment completion. That interval --- the settlement window --- exposed both parties to the potential loss of full principal. The systemic implications of bilateral interbank settlement exposures are formalised by Freixas, Parigi and Rochet (Freixas et al., 2000). Banks responded the only way they could: by holding a reserve against it. The reserve was not a regulatory requirement. It was a rational balance sheet response to the possibility of a repeat Herstatt. It sat on every cross-border banking balance sheet for the next two decades, and it constrained the volume of cross-border activity a given stock of money could support.

The formal structure of that constraint is straightforward. The theoretical treatment of settlement as a determinant of financial intermediation depth traces to Gorton and Pennacchi and Kahn and Roberds (Gorton \& Pennacchi, 1990; Kahn \& Roberds, 2009). A representative bank allocates its total balance sheet capacity B between a settlement buffer S --- required to cover potential settlement exposures --- and cross-border loans L, which generate the financial intermediation measured by inside money elasticity. The budget constraint is

\begin{equation}
S + L = B
\end{equation}

Settlement technology \ensuremath{\theta} \ensuremath{\in} {[}0,1{]} determines the required buffer, where \ensuremath{\theta} = 0 represents no settlement infrastructure and \ensuremath{\theta} = 1 represents perfect atomic settlement with zero exposure window. The buffer requirement is proportional to the outstanding loan book, scaled by the fraction of each loan that remains exposed during settlement:

\begin{equation}
S(\theta) = \alpha\,(1-\theta)\cdot L
\end{equation}

where \ensuremath{\alpha} \textgreater{} 0 is the balance sheet efficiency parameter. Substituting into the budget constraint gives equilibrium loans

\begin{equation}
L(\theta) = \frac{B}{1 + \alpha(1-\theta)}
\end{equation}

Inside money elasticity E \ensuremath{\equiv} L / M2:

\begin{equation}
E(\theta) = \frac{B}{M2 \cdot \bigl(1 + \alpha(1-\theta)\bigr)}
\end{equation}

The long-run comparative static is unambiguous: \ensuremath{\partial}E/\ensuremath{\partial}\ensuremath{\theta} \textgreater{} 0 for all \ensuremath{\theta}. Every improvement in settlement quality raises cross-border banking capacity. This is the prediction of the liberation phase.

But the model as stated cannot predict the compliance valley --- the post-2007 phase in which mandatory collateral requirements compress elasticity even as settlement quality continues to improve. If more settlement quality always raises elasticity, the 40 percent compression between 2007 and 2024 --- a period during which settlement quality continued improving --- is a contradiction. Resolving that contradiction requires decomposing the balance sheet efficiency parameter \ensuremath{\alpha} into the three distinct components that settlement reform history actually addressed.

\begin{equation}
\alpha = \alpha_{\mathrm{PAR}}(\theta_{\mathrm{PAR}}) + \alpha_{\mathrm{LPF}}(\theta_{\mathrm{LPF}}) + \alpha_{\mathrm{COL}}(\theta_{\mathrm{COL}})
\end{equation}

\subsection{II.B Three Channels, Three Institutional Mechanisms}
The balance sheet efficiency parameter \ensuremath{\alpha} is not a single number. It has three components, each targeted by a distinct generation of settlement reform, each operating through a different treasury function at a different time horizon. The additive structure follows from institutional separability: PAR exposure sits at the T+1 to T+2 settlement horizon, LPF prefunding sits at the intraday T+0 horizon, and COL margin sits at the multi-month derivative tenor horizon. These three pools do not substitute for each other. A euro of initial margin posted to a central counterparty does not reduce the intraday liquidity required to settle an FX leg, and neither reduces the overnight exposure to an unsettled cross-border payment. The channels are structurally distinct. This is why equation (5) aggregates them additively rather than multiplicatively: a multiplicative form would let any single fully-covered channel collapse the product and drive the inside-money ratio to zero, which the panel contradicts, since economies with high PAR and LPF coverage but low COL coverage retain substantial cross-border claims.

\textbf{The PAR channel: closing the Herstatt window.}

\ensuremath{\alpha}\_PAR is the principal-at-risk component --- the fraction of the loan book reserved against the potential loss of full trade value if the counterparty defaults before settlement completion. Delivery-versus-payment mechanics make the exchange of securities and cash simultaneous, eliminating the interval during which principal can be lost. Payment-versus-payment through CLS eliminates the same interval for foreign exchange transactions. Settlement finality legislation makes completed payments legally irrevocable under any insolvency scenario. As \ensuremath{\theta}\_PAR \ensuremath{\rightarrow} 1, \ensuremath{\alpha}\_PAR \ensuremath{\rightarrow} 0: the Herstatt buffer approaches zero.

\textbf{The LPF channel: eliminating the prefunding cost.}

\ensuremath{\alpha}\_LPF is the liquidity prefunding component --- the fraction of the loan book reserved to cover intraday gross payment obligations before multilateral netting provides offsetting relief. Real-time gross settlement systems with liquidity-saving mechanisms allow banks to offset incoming against outgoing payments in real time, compressing the intraday funding requirement. CLS applies the same logic to foreign exchange: by multilaterally netting FX obligations for all member banks simultaneously, it replaces bilateral gross settlement with a single net funding requirement per currency. T2S applies it to securities: by centralising settlement across European CSDs on a single platform, it allows collateral to serve multiple settlement obligations simultaneously. As \ensuremath{\theta}\_LPF \ensuremath{\rightarrow} 1, \ensuremath{\alpha}\_LPF \ensuremath{\rightarrow} 0.

\textbf{The COL channel: the asymmetric reform.}

\ensuremath{\alpha}\_COL is the collateral immobilisation component --- the fraction of the balance sheet held as mandatory initial margin in CCP collateral pools post-GFC. Here the direction reverses. PAR and LPF reforms reduced \ensuremath{\alpha} --- they shrank existing buffers. Mandatory collateral requirements did not reduce an existing buffer; they created a new mandatory one where none previously existed. From the balance sheet identity, \ensuremath{\partial}E/\ensuremath{\partial}\ensuremath{\alpha}\_COL \textless{} 0: each unit of mandatory margin compresses inside money elasticity. And unlike PAR and LPF, which were eroding reserves from a historically large starting point, this requirement imposed mandatory margin from a near-zero base --- entering the denominator with full force, unattenuated by any prior level of the buffer. This asymmetry is the precise mechanism behind the 2007--2024 inversion. What the framework predicts is an S-curve in SMI space with a compliance valley on the right-hand arm; what Sections IV and VI document is that this prediction holds in the panel, with two turning points, a channel-specific decomposition, and a sign pattern on COL that reverses the sign on PAR and LPF.

\subsection{II.C The Atomic Settlement Proof}
The initial margin requirement under post-GFC mandatory collateral requirements covers potential future exposure during the settlement window --- the period between the moment a trade is agreed and the moment its cash and securities legs are finally and irrevocably settled. The requirement is not arbitrary. It reflects the precise lesson of the Herstatt collapse: if a counterparty defaults after one leg has been transferred but before the other has settled, the surviving party faces the full principal loss of the unsettled leg. Initial margin is the pre-positioned collateral that covers that loss. Its size is calibrated to the maximum mark-to-market movement plausible over the duration of the window. The window duration is not a constant. It is an institutional variable --- determined by settlement technology, legal finality rules, and market convention --- and it has been shrinking throughout the panel\textquotesingle s 32-year sample as each generation of settlement reform has compressed the interval between agreement and completion.

\begin{equation}
\text{Initial margin requirement} = f(\text{window duration},\ \text{position volatility})
\end{equation}

Atomic delivery-versus-payment settlement compresses the window to zero by construction. In an atomic transaction, the transfer of the security and the transfer of the cash execute as a single indivisible event on a distributed ledger. There is no interval during which one leg has completed and the other has not. The transaction either completes in full, simultaneously, or it does not complete at all. The Herstatt exposure does not exist because the interval does not exist. The ECB\textquotesingle s Pontes architecture enforces atomic coupling between the DLT transaction leg and the TARGET cash settlement leg through Hash Time-Locked Contract mechanisms, delivering functional all-or-none settlement at the operational level.

\textbf{Proposition 1.} \emph{For any trade covered by a collateral requirement and settled atomically through a composable DLT protocol, the initial margin requirement converges to zero as the settlement window duration converges to zero. Formally: if f(\ensuremath{\tau}, \ensuremath{\sigma}) is a continuous margin function with f(0, \ensuremath{\sigma}) = 0 for all \ensuremath{\sigma} \ensuremath{\geq} 0, and atomic DLT settlement sets \ensuremath{\tau} = 0, then Initial margin \ensuremath{\equiv} f(0, \ensuremath{\sigma}) = 0.}

\textbf{Proof.} The margin requirement f(\ensuremath{\tau}, \ensuremath{\sigma}) is proportional to maximum mark-to-market loss over {[}0, \ensuremath{\tau}{]}. When \ensuremath{\tau} = 0 the exposure interval is empty; there is no temporal gap during which one leg completes and the other does not. The potential future exposure over a zero-duration window is zero for any position volatility \ensuremath{\sigma}. Substituting \ensuremath{\tau} = 0 into f gives f(0, \ensuremath{\sigma}) = 0.

Proposition 1 has one critical qualifier. It applies to covered trades --- those for which atomic settlement is the actual settlement mechanism for both counterparties. A trade between a German bank and a French bank is covered by Proposition 1 only if both banks are settling on atomic platforms and the atomic settlement of that specific trade is operationally live. A German bank settling atomically while its French counterparty settles conventionally does not satisfy the qualifier. The bilateral settlement window remains non-zero at the French end. The margin rationale is unaffected for that trade. Proposition 1 requires both counterparties on atomic platforms, not one. The efficiency gain from atomic settlement is therefore not a function of a single economy\textquotesingle s technology adoption. It is a function of bilateral pair completion --- a network property, not a node property.

\subsection{II.D The Network Efficiency Variable}

Proposition 1's qualifier has a formal expression. The Network Efficiency Variable (NEV) measures the efficiency available from atomic settlement as the volume-weighted sum of bilateral pairs across economies with atomic settlement live. For volume shares w\_i with \ensuremath{\Sigma}w\_i = 1 and interoperable economies set S:

\begin{equation}
\mathrm{NEV}(S) = \sum_{i \in S}\ \sum_{j \in S,\, j \neq i} w_i \cdot w_j
\end{equation}

Equation (7) is Metcalfe's network value formula (Economides, 1996; Farrell \& Saloner, 1985; Katz \& Shapiro, 1985; Metcalfe, 2013) with two adjustments. First, pairs are weighted by settlement volume rather than counted equally, addressing the Briscoe, Odlyzko and Tilly critique that not all bilateral connections carry equal weight; under uniform weights, equation (7) collapses to the canonical n(n-1)/2 form, while concentrated weights deliver the n\ensuremath{\cdot}log(n) form Briscoe et al. propose (Briscoe et al., 2006). Zhang, Liu and Xu confirm this volume-weighted specification matches observed network adoption on Tencent and Facebook data (Zhang et al., 2015). The empirical BIS Locational Banking Statistics distribution sits between the uniform and concentrated extremes.

The formula has a sharp corollary. For a single economy adopting in isolation, S = \{i\}, the sum has no terms and NEV = 0. The result is mechanical: atomic settlement requires both counterparties on atomic platforms, and a one-economy set contains no pairs. Solo adoption produces zero efficiency regardless of how large the economy is or how good its technology.

Switzerland tests this prediction directly. The SIX Digital Exchange brought atomic delivery-versus-payment live in 2021. The synthetic control test in Section VII.B finds no efficiency effect on Swiss inside money elasticity over 2022--2024. With zero bilateral atomic pairs in the panel, NEV = 0 for Switzerland alone, exactly as the formula predicts. Proposition 1 is confirmed in its qualifier.

The NEV formula maps onto the BIS three-layer connectivity taxonomy (BIS, 2023). Fragmentation is the pre-network state of isolated nodes (NEV = 0). Interoperability is bilateral pairs through bridges and gateways --- the case equation (7) describes directly. Composability extends the formula beyond pairwise to multi-asset settlement clusters within a single instruction. Section VII applies these three layers to the Eurosystem's Pontes and Appia, calibrating the Pontes forecast against the T2S coordinated regulatory migration to derive a 2027--2032 efficiency recovery corridor.

\section{III. Data and the Settlement Modernisation Index}

\subsection{III.A The 809 Events: What We Coded and Why}
No existing index measures cumulative settlement infrastructure development at the country level over time. The construction protocol follows Abiad, Detragiache and Tressel (2010), who translate discrete financial-liberalisation events across seven policy dimensions into a continuous country-panel index; what changes here is the domain, since the events coded are regulatory frameworks, technology go-lives and market practice thresholds rather than capital account opening or interest rate deregulation. We construct the Settlement Modernisation Index (SMI) using discrete institutional reforms coded from primary legal sources, assigned to economic channels and adoption phases, and accumulated into a continuous panel measure. The panel window is 1993 to 2024, defined by the availability of panel-wide BIS Locational Banking Statistics on cross-border claims; the index encodes 809 reform events across 24 advanced economies. The construction proceeds in five steps.

\textbf{Step 1: Taxonomy.} The twelve categories (T-01 through T-12) are derived in three sub-steps. First, we adopt the three economic channels of Section II --- principal-at-risk, liquidity and payment finality, and collateral immobilisation --- as the inclusion criterion: an institutional reform is in scope if and only if it operates on at least one of the three channels through the balance-sheet mechanism in equation (5). Second, we enumerate the candidate reform population from primary sources as detailed in Table 1. Third, we cluster the enumerated reforms by mechanism: two reforms collapse into a single category if they affect the same channel through the same balance-sheet mechanism, regardless of jurisdiction or instrument type --- so, for example, the EU's CSDR settlement-cycle compression and SEC Rule 15c6-1 are both T-01 (Settlement Cycle Compression), because the balance-sheet effect is the same shortened exposure window. The taxonomy covers all reforms identified in the source catalogue that affect inside money elasticity through equation (5). Two structurally absent phases are recorded: T-04 has no MK phase because settlement finality is a legal property rather than a market practice, and T-08 (International Standards Adoption) has no TH phase because standards are adopted by national authorities rather than deployed as technical infrastructure. Net of these structural exclusions, the master taxonomy contains 22 PAR-relevant event-phase slots, 31 LPF-relevant slots, and 20 COL-relevant slots, defining the universal ceilings against which all countries are normalised in Step 5.

\textbf{Step 2: Country-specific applicability and date determination.} For each country, we identify which of the 12 event types are applicable given that country's regulatory regime, legal system, and market structure. Date determination depends on the legal nature of the instrument that establishes the reform, with four major cases distinguished. \emph{EU Regulations} --- including EMIR (Regulation 648/2012), CSDR (Regulation 909/2014), and the DLT Pilot Regime (Regulation 2022/858) --- are directly applicable in all Member States from the date of entry into force, without national transposition; we therefore assign the same RF year to all EU members for these instruments, equal to the entry-into-force date specified in the Regulation. \emph{EU Directives}, including the Settlement Finality Directive 98/26/EC and the Financial Collateral Directive 2002/47/EC, require national transposition into domestic law within a stated deadline; we assign the RF year for each Member State equal to the year of the country's national transposing act, identified from EU notification records and confirmed against national legal gazettes. \emph{ECB-operated platforms}, principally TARGET2 and TARGET2-Securities, are not legal instruments under Article 288 TFEU but operate under ECB Decisions and Framework Agreements with participating central securities depositories; we assign the RF year as the year of the relevant ECB framework decision (2012 for T2S) and the TH year as the country's wave-migration date. \emph{Non-EU countries} --- the United Kingdom post-Brexit, Switzerland, Canada, Norway, the United States, Australia, Japan, South Korea, Singapore, and Hong Kong --- require a nationally applicable equivalent law, regulation, or central bank rule before any event is coded as applicable, with the RF year set equal to the date of the national instrument.

\textbf{Step 3: Phase dating.} For each applicable event we identify three adoption dates from primary sources. The RF year is the year the regulatory framework enabling the reform was enacted, determined according to the rules in Step 2. The TH year is the year the technology or infrastructure became operationally live. The MK year is the year the reform reached systemic market practice, defined as the year in which the reform became the routine default across all eligible transactions in the relevant market, evidenced by primary-source confirmation that subsequent settlement activity occurred under the reformed regime. MK dates are determined mainly from BIS CPMI annual payment statistics, ECB TARGET2 and T2S volume reports, central bank payment-system oversight reports, and CCP disclosure documents; where volume statistics are not available, the MK year is dated from the year in which the relevant BIS CPMI or FSB peer-review report first characterises the reform as embedded in market practice. Where a phase has not been completed as of December 2024, it is left uncoded. Reforms with adoption dates before the panel window are preserved at their historical year and enter the index at t\ensuremath{_{0}} = 1993, so that each country's index value in the first panel year already reflects its accumulated settlement infrastructure from prior decades; sixty-three of the 809 events (7.8 percent) have adoption years before 1993 and are accounted for through this t\ensuremath{_{0}} entry mechanism, while the remainder enter in their actual completion year.

\textbf{Step 4: Scoring.} Each event receives a raw score equal to the product of its channel-breadth score (L1 = channel count divided by 3) and its phase-depth score (L2 = phases complete divided by 3), yielding a per-event score on the interval {[}1/9, 1{]}. Channel scores are the sum of per-event scores for applicable, completed events, with maximum possible values under the master taxonomy.

\textbf{Step 5: Normalisation.} We normalise each channel score by a fixed universal ceiling --- the total score achievable if a country had completed all three phases of every reform event in the master taxonomy. The universal denominators are derived from the maximum number of applicable event-phase slots per channel under the taxonomy fixed in Step 1, net of the two structurally absent phases identified in that step. This places each channel on the unit interval {[}0, 1{]}, where unity represents full completion of every reform event in every phase. The composite SMI is the arithmetic mean of the three universal-ceiling-normalised channel scores. Universal-ceiling normalisation is preferred over three alternatives on theoretical grounds. Normalising each channel by the country's own maximum possible score under its applicable-event subset (V1) mechanically rescales every country's terminal value toward a common plateau, eliminating the cross-country level variation that the efficiency framework in Section II requires the index to capture. Normalising by a single global benchmark country such as the United States (V2) absorbs cross-country variation in applicable events into country fixed effects rather than identifying it through the index. Normalising by the annual panel maximum (V3) destroys the S-curve turning-point structure by placing \ensuremath{\tau}\ensuremath{_{1}}* outside {[}0, 1{]}. Universal-ceiling normalisation alone preserves both the within-country progression that identifies the cubic specification and the across-country level variation that allows comparisons of regulatory coverage.

The continuous, channel-decomposed, universal-ceiling-normalised index defined in Steps 1--5 is the appropriate main explanatory variable for the empirical analysis that follows in Section IV.

\textbf{Table 1 --- SMI Taxonomy: Twelve Reform Categories, Three Economic Channels, and Three Adoption Phases}

\begin{longtable}[]{@{}
  >{\raggedright\arraybackslash}p{(\columnwidth - 14\tabcolsep) * \real{0.0753}}
  >{\raggedright\arraybackslash}p{(\columnwidth - 14\tabcolsep) * \real{0.1971}}
  >{\raggedright\arraybackslash}p{(\columnwidth - 14\tabcolsep) * \real{0.0884}}
  >{\raggedright\arraybackslash}p{(\columnwidth - 14\tabcolsep) * \real{0.0757}}
  >{\raggedright\arraybackslash}p{(\columnwidth - 14\tabcolsep) * \real{0.0606}}
  >{\raggedright\arraybackslash}p{(\columnwidth - 14\tabcolsep) * \real{0.0936}}
  >{\raggedright\arraybackslash}p{(\columnwidth - 14\tabcolsep) * \real{0.1819}}
  >{\raggedright\arraybackslash}p{(\columnwidth - 14\tabcolsep) * \real{0.2274}}@{}}
\toprule\noalign{}
\begin{minipage}[b]{\linewidth}\raggedright
\textbf{Code}
\end{minipage} & \begin{minipage}[b]{\linewidth}\raggedright
\textbf{Category}
\end{minipage} & \begin{minipage}[b]{\linewidth}\raggedright
\textbf{PAR}
\end{minipage} & \begin{minipage}[b]{\linewidth}\raggedright
\textbf{LPF}
\end{minipage} & \begin{minipage}[b]{\linewidth}\raggedright
\textbf{COL}
\end{minipage} & \begin{minipage}[b]{\linewidth}\raggedright
\textbf{Active channels}
\end{minipage} & \begin{minipage}[b]{\linewidth}\raggedright
\textbf{Main Primary sources}
\end{minipage} & \begin{minipage}[b]{\linewidth}\raggedright
\textbf{Economic mechanism and balance-sheet effect}
\end{minipage} \\
\midrule\noalign{}
\endhead
\bottomrule\noalign{}
\endlastfoot
\multicolumn{8}{@{}l@{}}{%
\textbf{Panel A --- Twelve Taxonomy Categories}} \\
\textbf{T-01} & Settlement Cycle Compression & --- & \ensuremath{\checkmark} & \ensuremath{\checkmark} & LPF, COL & EU CSDR Reg. 909/2014 Art. 5(2); SEC Rule 15c6-1; BIS CPSS Red Book Vol. III ch. 2 & Compresses the rolling settlement exposure window, reducing gross overnight funding requirements and mandatory CCP margin on open positions \\
\textbf{T-02} & Delivery-versus-Payment & \ensuremath{\checkmark} & --- & --- & PAR & BIS CPSS Delivery versus Payment in Securities Settlement Systems (1992); CSDR Reg. 909/2014 Art. 6; CPMI-IOSCO PFMI Principle 12 & Eliminates bilateral principal exposure by making security and cash transfers simultaneous and indivisible \\
\textbf{T-03} & Payment-versus-Payment / CLS & \ensuremath{\checkmark} & \ensuremath{\checkmark} & --- & PAR, LPF & BIS CPSS Settlement Risk in Foreign Exchange Transactions (1996, 2008); CLS Bank International Annual Reports; BIS Triennial Central Bank Survey & Applies DvP logic to FX: multilateral netting eliminates gross FX settlement exposure and compresses intraday funding \\
\textbf{T-04} & Settlement Finality Legislation & \ensuremath{\checkmark} & \ensuremath{\checkmark} & --- & PAR, LPF & EU Settlement Finality Directive 98/26/EC; UNIDROIT 2009 Convention; national finality acts & Makes completed payments legally irrevocable under insolvency. Self-executing on enactment --- no MK phase required \\
\textbf{T-05} & Securities Dematerialisation & \ensuremath{\checkmark} & \ensuremath{\checkmark} & --- & PAR, LPF & National CSD legislation; BIS CPSS Red Book Vol. III ch. 2; ECSDA implementation guides & Eliminates physical certificate risk and enables electronic DvP; widest historical span in the panel \\
\textbf{T-06} & CSD Platform Migration (T2S) & --- & \ensuremath{\checkmark} & \ensuremath{\checkmark} & LPF, COL & EU CSDR Reg. 909/2014; ECB T2S Framework Agreement 2012; ECB T2S Migration Reports 2015--2017 & Centralises CSD settlement enabling cross-border collateral mobility and multi-obligation netting \\
\textbf{T-07} & CCP and Mandatory Clearing & \ensuremath{\checkmark} & \ensuremath{\checkmark} & \ensuremath{\checkmark} & PAR, LPF, COL & EU EMIR Reg. 648/2012 Art. 4; CFTC Dodd-Frank Title VII; CPMI-IOSCO PFMI Principle 14; FSB OTC Derivatives Reform Progress Reports & Primary driver of the compliance valley: imposes mandatory initial margin from near-zero base with full denominator force. \\
\textbf{T-08} & International Standards Adoption & \ensuremath{\checkmark} & \ensuremath{\checkmark} & \ensuremath{\checkmark} & PAR, LPF, COL & CPMI-IOSCO PFMI 2012; FSB Standards Implementation Monitoring; IMF FSAP country reports & Standards-direct: no technology phase. Efficiency follows regulatory and market practice embedding of PFMI compliance \\
\textbf{T-09} & Financial Collateral Framework & --- & \ensuremath{\checkmark} & \ensuremath{\checkmark} & LPF, COL & EU Financial Collateral Directive 2002/47/EC; UNIDROIT Convention on Intermediated Securities 2009; national collateral law & Establishes cross-border collateral mobility and rehypothecation rights, enabling netting across jurisdictions \\
\textbf{T-10} & RTGS / Hybrid Payment System & \ensuremath{\checkmark} & \ensuremath{\checkmark} & --- & PAR, LPF & BIS CPSS Red Book Vol. I--III ch. 1; ECB TARGET2 Information Guides; BoE CHAPS Reference Manual & Each RTGS generation compresses intraday gross exposure and enables real-time finality \\
\textbf{T-11} & Settlement Discipline Regime & --- & \ensuremath{\checkmark} & \ensuremath{\checkmark} & LPF, COL & EU CSDR Reg. 909/2014 Art. 7; Settlement Discipline Reg. (EU) 2018/1229; ECB Settlement Discipline Guidance & Settlement fail penalties and mandatory buy-in obligations under CSDR Art. 7 impose operational compliance costs through LPF efficiency and COL margin channels \\
\textbf{T-12} & DLT and Tokenisation Framework & \ensuremath{\checkmark} & \ensuremath{\checkmark} & \ensuremath{\checkmark} & PAR, LPF, COL & EU DLT Pilot Reg. 2022/858; Swiss DLT Act / FinIA-A 2021; ECB Pontes framework decision 2026; SIX SDX Annual Reports 2022--2024 & Closes settlement window to zero by construction; eliminates COL margin rationale on covered trades. Efficiency requires bilateral pairs \\
\multicolumn{8}{@{}l@{}}{%
\textbf{Panel B --- Three Adoption Phases}} \\
\textbf{Phase} & \textbf{Full name} & \multicolumn{6}{@{}>{\raggedright\arraybackslash}p{0.70\columnwidth}@{}}{%
\textbf{Definition and efficiency role}} \\
\textbf{RF} & Regulatory Framework Enacted & \multicolumn{6}{@{}>{\raggedright\arraybackslash}p{0.70\columnwidth}@{}}{%
Primary legislation or binding regulation published in official journal. Establishes legal obligation and timeline.} \\
\textbf{TH} & Technology Operationally Live & \multicolumn{6}{@{}>{\raggedright\arraybackslash}p{0.70\columnwidth}@{}}{%
Settlement infrastructure switched on and available for use. T-08 has no TH entries (standards require no technology).} \\
\textbf{MK} & Market Practice Embedded & \multicolumn{6}{@{}>{\raggedright\arraybackslash}p{0.70\columnwidth}@{}}{%
The reform has become the routine default across all eligible transactions in the relevant market. This is the efficiency event: the balance sheet responds to what actually occurs in settlement, not to what is regulatorily mandated or technically available. T-04 has no MK entries (finality is self-executing on enactment).} \\
\end{longtable}

The distribution of the 809 events across time and across channels is not uniform, and its unevenness is itself part of the measurement. Three patterns in particular shape the analysis that follows. First, the twelve taxonomy categories concentrate in different institutional eras: PAR-dominant categories (T-02 Delivery-versus-Payment, T-04 Settlement Finality Legislation) cluster in the pre-2007 liberation phase, with 65 and 90 percent of their events respectively completed before 2007. LPF-and-COL categories (T-01 Settlement Cycle Compression, T-06 CSD Platform Migration, T-11 Settlement Discipline Regime) concentrate in the post-2012 mandatory clearing era, with 85, 69, and 94 percent of their events respectively dated post-GFC collateral requirement entry into force. The timing of category activation mirrors the timing of the institutional eras the Section IV estimates recover from the data. Second, T-12 (DLT and Tokenisation Framework) has zero events before 2019 by construction and 2 events thereafter --- reflecting the nascent state of atomic DLT adoption in the 1993--2024 panel window, which is what makes the forward prediction in Section VII a genuine out-of-sample exercise rather than an in-sample fit. Third, the event counts differ across categories (from 48 in T-04 and T-08 to 110 in T-10) not because of taxonomic inflation but because the underlying reform mechanisms differ in their institutional complexity: RTGS modernisation (T-10) has had four to five generational iterations per economy across the sample, while settlement finality legislation (T-04) is typically a single enactment event per jurisdiction. Figure 1 visualises the full distribution.

\textbf{Figure 1: Settlement Reform Event Density by Era}

\begin{figure}[H]
\centering
\includegraphics[width=\linewidth,height=3.3in,keepaspectratio]{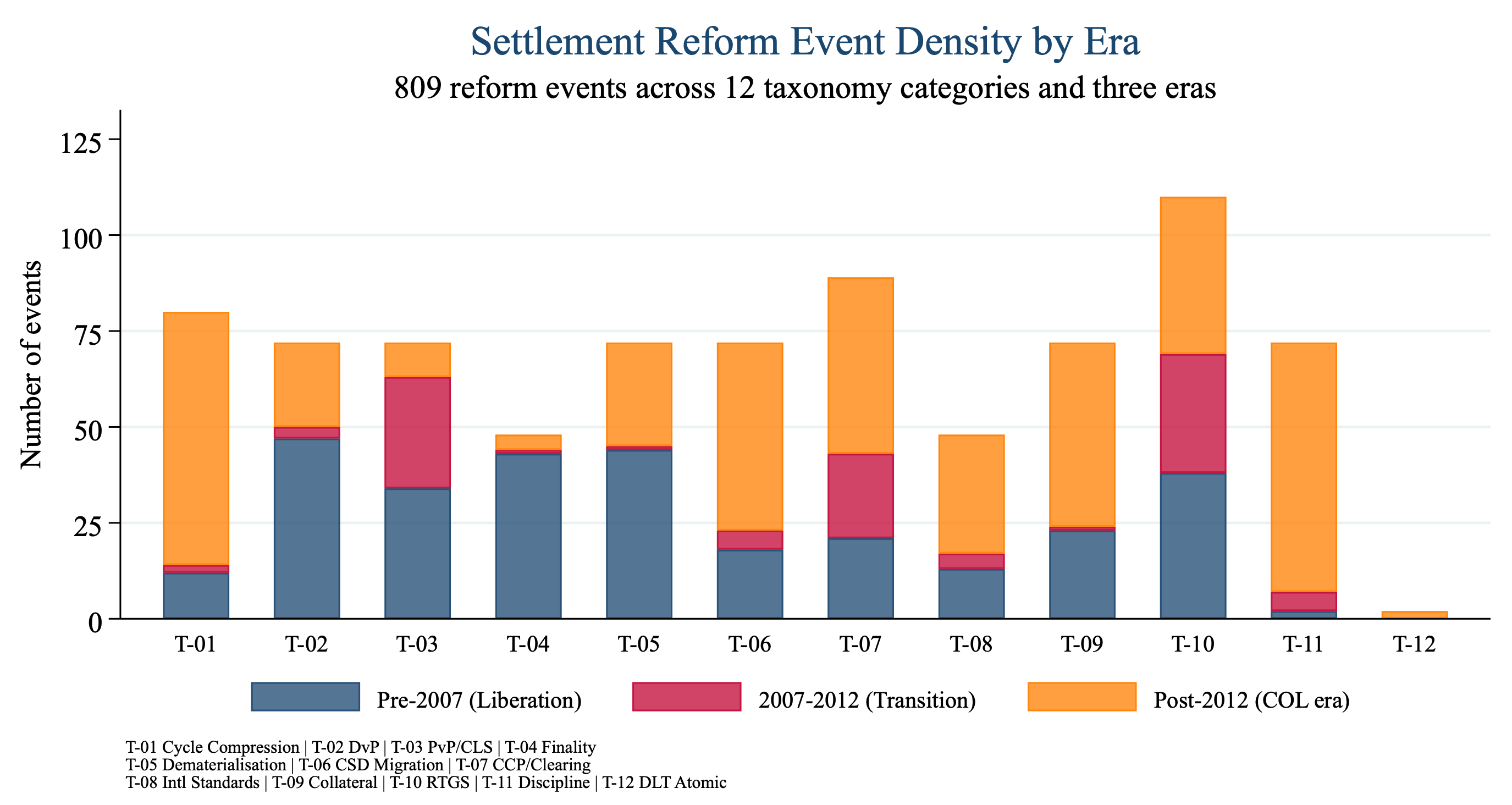}
\end{figure}

\subsection{III.B The Dependent Variable: Inside Money Elasticity}
The dependent variable throughout the analysis is the natural logarithm of the ratio of cross-border bank claims to broad money supply, ln(CB/M2), measured annually for each panel economy. Cross-border bank claims are drawn from the BIS Locational Banking Statistics. Broad money supply is drawn from a combination of the World Bank World Development Indicators for five non-eurozone economies (Australia, Denmark, Hong Kong, Japan, Norway), Bloomberg terminal data for fourteen economies with Bloomberg-available M2 series, and national central bank publications for the five remaining eurozone economies where Bloomberg coverage is incomplete (Austria, Belgium, Luxembourg, the Netherlands, Portugal).

The ratio captures the volume of cross-border financial intermediation that a given stock of inside money can support. The ratio belongs to a family of measures of cross-border financial integration. Chinn and Ito (2006) capture openness de jure, from the provisions of capital control regimes; Lane and Milesi-Ferretti (2007) capture it de facto, through stocks of external assets and liabilities. Inside money elasticity differs from both in locating integration on the banking side of the balance sheet, where the settlement constraint of Section II binds. That is the property required here: a measure sensitive to whether banks can extend cross-border claims against a given stock of inside money, rather than to whether capital is legally free to move. To the extent that settlement infrastructure improvements release balance-sheet capacity --- by reducing principal-at-risk buffers, compressing intraday prefunding requirements, or eliminating mandatory margin obligations --- the CB/M2 ratio rises and efficiency improves. To the extent that settlement reform imposes new mandatory obligations --- as EMIR\textquotesingle s initial margin requirement does --- the ratio falls and efficiency compresses. The ratio is not a price. It is a quantity: how much cross-border banking the monetary base currently sustains.

The logarithmic transformation is required for the cubic polynomial specification to have an economically meaningful marginal effect interpretation and stabilises the variance across the wide cross-sectional range --- which spans from 0.057 times to 7.604 times, a range of two orders of magnitude that would create severe heteroscedasticity in levels.

The panel is unbalanced. Of the 24 economies, 10 have data beginning in 1993 --- Canada, Switzerland, Denmark, the United Kingdom, Hong Kong, Japan, Norway, Sweden, Singapore, and the United States. Fourteen economies have shorter series: the eurozone economies and several others whose cross-border banking statistics begin in 1999, with Portugal and Australia beginning in 2001, Belgium and Greece and Ireland in 2002--2003, and Korea beginning only in 2005. The panel contains 663 country-year observations in total. All specifications include country fixed effects, year fixed effects and macroeconomic controls, as described in Section III.C.

\subsection{III.C Control Variables}
The primary regression includes two time-varying controls: real GDP growth and the central bank policy rate, sourced from Bloomberg and FRED with country-specific series for the United States, Singapore, Hong Kong, and the eurozone, where the ECB main refinancing rate applies from 1999 onwards. Both controls absorb within-country business cycle variation that might otherwise confound the reform signal. Two additional panel-wide variables --- the CBOE Volatility Index (VIX) and the US Federal Funds Rate --- are reported in the descriptive statistics and enter the robustness specification in Table 5, but are excluded from the primary regression because their variation is common to the panel and absorbed by the year fixed effects.

\subsection{III.D Descriptive Statistics}
The 809 SMI events are not the unit of regression analysis. The regressions use country-year observations on a 1993--2024 unbalanced panel: N = 663 across 24 countries and 32 calendar years. Country-year missingness arises from staggered country entry into the BIS Locational Banking Statistics panel --- Table A2 cross-border claims series begin in different years for different reporters --- and from incomplete dependent-variable coverage where either the numerator (cross-border claims) or the denominator (broad money) is missing for a given country-year.

\textbf{Table 2 --- Descriptive Statistics, Panel Variables 1993--2024}

\begin{longtable}[]{@{}
  >{\raggedright\arraybackslash}p{(\columnwidth - 10\tabcolsep) * \real{0.1667}}
  >{\raggedright\arraybackslash}p{(\columnwidth - 10\tabcolsep) * \real{0.1667}}
  >{\raggedright\arraybackslash}p{(\columnwidth - 10\tabcolsep) * \real{0.1667}}
  >{\raggedright\arraybackslash}p{(\columnwidth - 10\tabcolsep) * \real{0.1667}}
  >{\raggedright\arraybackslash}p{(\columnwidth - 10\tabcolsep) * \real{0.1667}}
  >{\raggedright\arraybackslash}p{(\columnwidth - 10\tabcolsep) * \real{0.1667}}@{}}
\toprule\noalign{}
\begin{minipage}[b]{\linewidth}\raggedright
\textbf{Variable}
\end{minipage} & \begin{minipage}[b]{\linewidth}\raggedright
\textbf{N}
\end{minipage} & \begin{minipage}[b]{\linewidth}\raggedright
\textbf{Mean}
\end{minipage} & \begin{minipage}[b]{\linewidth}\raggedright
\textbf{Std. Dev.}
\end{minipage} & \begin{minipage}[b]{\linewidth}\raggedright
\textbf{Min}
\end{minipage} & \begin{minipage}[b]{\linewidth}\raggedright
\textbf{Max}
\end{minipage} \\
\midrule\noalign{}
\endhead
\bottomrule\noalign{}
\endlastfoot
ln(CB/M2) & 663 & -0.411 & 0.983 & -2.872 & 2.029 \\
nSMI\_Total & 768 & 0.426 & 0.283 & 0.000 & 0.957 \\
nSMI\_PAR & 768 & 0.523 & 0.277 & 0.000 & 0.955 \\
nSMI\_LPF & 768 & 0.432 & 0.293 & 0.000 & 0.968 \\
nSMI\_COL & 768 & 0.323 & 0.292 & 0.000 & 0.950 \\
GDP growth (\%) & 768 & 2.315 & 3.112 & -10.940 & 24.624 \\
Policy rate (\%) & 768 & 2.569 & 2.534 & -1.267 & 14.983 \\
VIX & 768 & 19.514 & 5.913 & 11.090 & 32.695 \\
\end{longtable}

\subsection{III.E When Does Reform Deliver? The TH--MK Evidence}
The single most important distinction in the SMI methodology is the separation of the technology phase from the market practice phase. Settlement reforms are routinely described by their technology go-live dates: CLS launched in 2002, T2S went live in 2015. These dates are accurate for the technology. They are not the dates at which efficiency appeared.

The DvP evidence establishes this across the full 24-country panel. Delivery-versus-payment technology was operational in the Netherlands in 1963. Systemic market practice --- the universal, default, routinely expected settlement of every eligible transaction on DvP terms --- was not established until 1992, a gap of 29 years. During those 29 years, DvP infrastructure existed. The efficiency effect waited for market practice because the balance sheet responds to what actually happens in settlement, not to what is technically possible. The same gap appears across the panel with varying lengths: the United States had DvP technology operational in 1967 and market practice established in 1992 --- 25 years. Germany operational in 1969, market practice 1992 --- 23 years. Denmark operational in 1983, market practice 2015 --- 32 years. Finland operational in 1993, market practice 2015 --- 22 years. Japan achieved TH in 1994 and MK in 1995 --- a one-year gap. Belgium and Luxembourg both achieved TH and MK in the same year.

The CLS evidence replicates the pattern at the network level. CLS technology was live from September 2002, initially processing roughly 7 percent of global FX market volume --- technology available but not yet operating at the scale required to deliver its full efficiency benefit. In 2005--2006, CLS crossed 35 percent of global FX volume (CLS Bank International, 2024) and the network effect materialised. The efficiency event was not the 2002 TH. It was the 2005--2006 market practice consolidation at which CLS became the routine expectation of the FX market.

The T2S evidence provides the most precisely timed illustration in the sample. T2S Wave 1 in June 2015 connected Austria\textquotesingle s CSD to the T2S platform. Austria\textquotesingle s CB/M2 fell 20.3 percent from 2015 to 2017 --- continued compression from the ongoing compliance valley, with no detectable T2S efficiency contribution. T2S Wave 3 in September 2016 and 2017 brought Deutsche Börse, the French, Italian, Dutch, Spanish, Portuguese, Finnish, Greek, and Irish CSDs --- the securities settlement volume of the eight largest EU economies arriving on the platform across a 14-month window. Wave 1 delivered no efficiency gain because the network lacked the volume nodes. Wave 3 delivered the efficiency gain because the network acquired them simultaneously.

The DLT evidence is the most recent and most direct confirmation. Switzerland's SDX completed T-12 TH in 2021 --- the first economy in the panel to bring atomic DLT settlement technology operationally live. The synthetic control test in Section VII.B finds no detectable efficiency effect from 2022 onward. Switzerland brought atomic DLT settlement technology operationally live for a network with no bilateral counterparties. The TH--MK distinction applies with equal force to DLT: the efficiency event is not the technology; it is the moment at which the network reaches the scale at which bilateral pair netting generates real balance-sheet relief.

\subsection{III.F Index Validation}
The SMI\textquotesingle s three-phase structure generates a specific econometric prediction: no single phase sub-index should explain inside money elasticity as well as the composite index that requires all three phases to fire. If efficiency followed regulatory enactment alone, the RF sub-index would dominate. If efficiency followed technology deployment alone, the TH sub-index would dominate. If the composite produces a meaningful multiple of every single-phase alternative, the three-phase structure is the right encoding of how settlement reform affects balance sheets.

The validation runs the test directly. Regressing ln(CB/M2) on each phase sub-index (RF-only, TH-only, MK-only) with country and year fixed effects and the same macro controls as the primary specification yields within-R\ensuremath{^{2}} values that are uniformly small. The composite SMI yields within-R\ensuremath{^{2}} = 0.123, materially larger than the explanatory power of any individual phase sub-index. No individual phase carries the efficiency signal on its own; the interaction of the three phases carries it.\hspace{0pt}\hspace{0pt}\hspace{0pt}\hspace{0pt}\hspace{0pt}\hspace{0pt}\hspace{0pt}\hspace{0pt}\hspace{0pt}\hspace{0pt}\hspace{0pt}\hspace{0pt}\hspace{0pt}\hspace{0pt}\hspace{0pt}\hspace{0pt}The DvP, CLS, T2S, and DLT evidence in Section III.E provides the narrative counterpart to this result. efficiency events align with MK completions specifically: the Netherlands 1992 DvP efficiency moment (not the 1963 TH), the 2005--2006 CLS network crossing (not the 2002 TH), the T2S Wave 3 volume consolidation (not the Wave 1 TH). Switzerland\textquotesingle s SDX null confirms the same pattern in reverse: TH without bilateral pairs produces no efficiency effect, exactly as the framework predicts.

The composite SMI that encodes all three phases is therefore both the econometrically best-performing specification and the narratively correct one. Sections IV through VI use the composite index as the primary right-hand-side variable; Sections III.E and III.F together justify that choice against the single-phase alternatives.

\section{IV. The Cross-border wholesale Settlement Arc}

\subsection{IV.A The Arc, 1993--2024}
In 1993, the average advanced economy in the panel held wholesale cross-border bank claims equal to 1.27 times its domestic broad money supply. By 2007, that ratio had risen to 1.42 times --- 12 percent above the 1993 baseline. By 2024, it had fallen to 0.84 times --- 40 percent below the 2007 peak, and 33 percent below the 1993 starting point. The panel\textquotesingle s average bank today holds proportionally fewer cross-border claims relative to its money supply than it did in the year the single European market for financial services came into existence. Three decades of settlement reform produced a rise, a peak, a collapse, and the beginning of a recovery. That reform effects can reverse sign is not without precedent. Kaminsky and Schmukler (2008) document a short-run-pain, long-run-gain pattern in which liberalisation amplifies stock market cycles immediately and stabilises them over longer horizons. Two things separate that finding from this one. Their outcome is asset-price behaviour rather than intermediation capacity, so no balance-sheet mechanism is identified; and their non-monotonicity is a property of time since reform rather than of reform intensity, so no turning point exists to be located. The contribution here is the mechanism generating the reversal and the estimated location of the two turning points it implies. Figure 2 plots the panel-average trajectory.

\textbf{Figure 2: The Settlement Arc, 1993-2024}

\begin{figure}[H]
\centering
\includegraphics[width=\linewidth,height=3.3in,keepaspectratio]{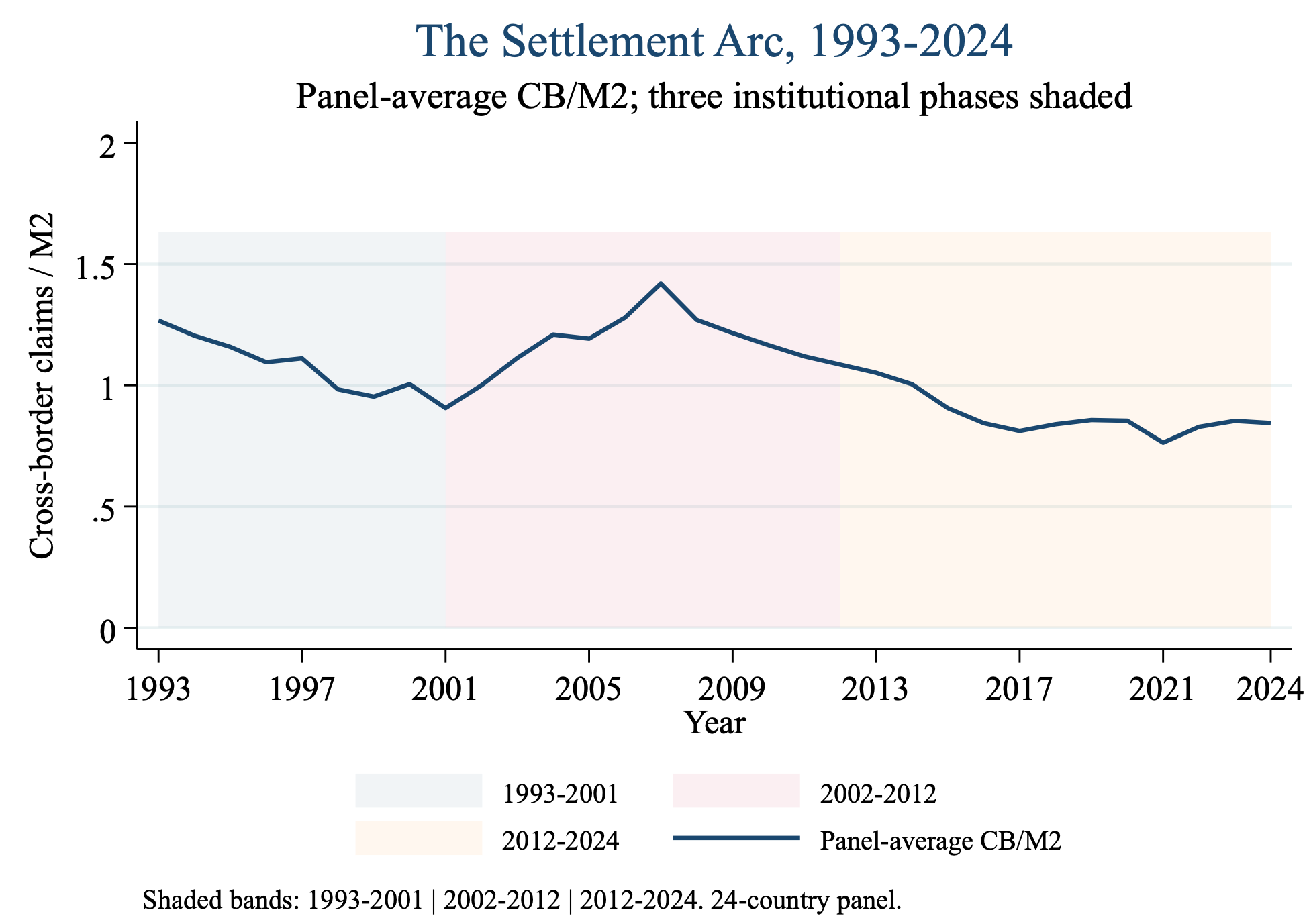}
\end{figure}

The first feature of the arc is its scale. The 40 percent compression from peak to 2024 is not a marginal adjustment. In absolute dollar terms, the mean per-country cross-border claims witnessed in the panel grew from \$378 billion in 1993 to \$1,503 billion in 2024--- a 298 percent increase in per-country average size. Banks did more cross-border business in 2024 than at any prior point in the sample. The ratio fell not because banks retreated from cross-border activity but mainly because a single mandatory clearing regulation imposed a new permanent balance-sheet obligation that absorbed capacity faster than three channels of settlement reform could release it.

The second feature is its distribution. The panel average conceals a cross-section that spans two orders of magnitude. Luxembourg held cross-border claims equal to 7.6 times its broad money supply at the 2007 peak --- the highest ratio recorded in the panel across the full 32-year sample. Singapore reached 5.75 times in 1993. Ireland reached 3.58 times in 2007. At the other end, Korea\textquotesingle s measured CB/M2 ratio has never exceeded 0.11 times in any year for which data are available. The United States ended 2024 at 0.12 times --- the second lowest in the panel, 56 percent below its own 2010 peak. The falls from individual country peaks are more striking than the panel average: Luxembourg fell 75 percent from its 2007 peak to 2024, Switzerland fell 63 percent, and Belgium fell 66 percent.

The third feature is the timing. The arc has three distinct phases with identifiable turning points --- an acceleration between 1993 and 2007 driven by principal-risk elimination and liquidity efficiency gains, an inversion from 2012 driven by mandatory collateral immobilisation, and a recovery phase that, as of 2024, only Switzerland has entered (SMI = 0.957, above \ensuremath{\tau}\ensuremath{_{2}}* = 0.934), while the remaining 23 economies have converged to the universal-ceiling plateau of 0.872 --- still within the compliance valley but at the threshold of recovery. Each phase corresponds to a dominant settlement reform channel. Section IV.B estimates the turning points and identifies where each country stood within the sequence.\hspace{0pt}\hspace{0pt}\hspace{0pt}\hspace{0pt}\hspace{0pt}\hspace{0pt}\hspace{0pt}\hspace{0pt}\hspace{0pt}\hspace{0pt}\hspace{0pt}\hspace{0pt}\hspace{0pt}\hspace{0pt}\hspace{0pt}\hspace{0pt}

\subsection{IV.B The S-Curve and Its Three Phases}
The relationship between settlement quality and inside money elasticity is estimated with a cubic polynomial. Three specifications are compared: linear, quadratic, and cubic. A Ramsey RESET test on the linear specification rejects linearity (F = 23.77, p \textless{} 0.001), confirming the presence of non-linear structure. The Bayesian model comparison produces a decisive result: the cubic specification improves the AIC by 54.59 points relative to the linear, while the quadratic improves by only 34.41 points. Figure 3 plots the implied marginal effect.

\textbf{Figure 3: Cubic S- Curve: Marginal Effect of SMI on ln(CB/M2)}

\begin{figure}[H]
\centering
\includegraphics[width=\linewidth,height=3.3in,keepaspectratio]{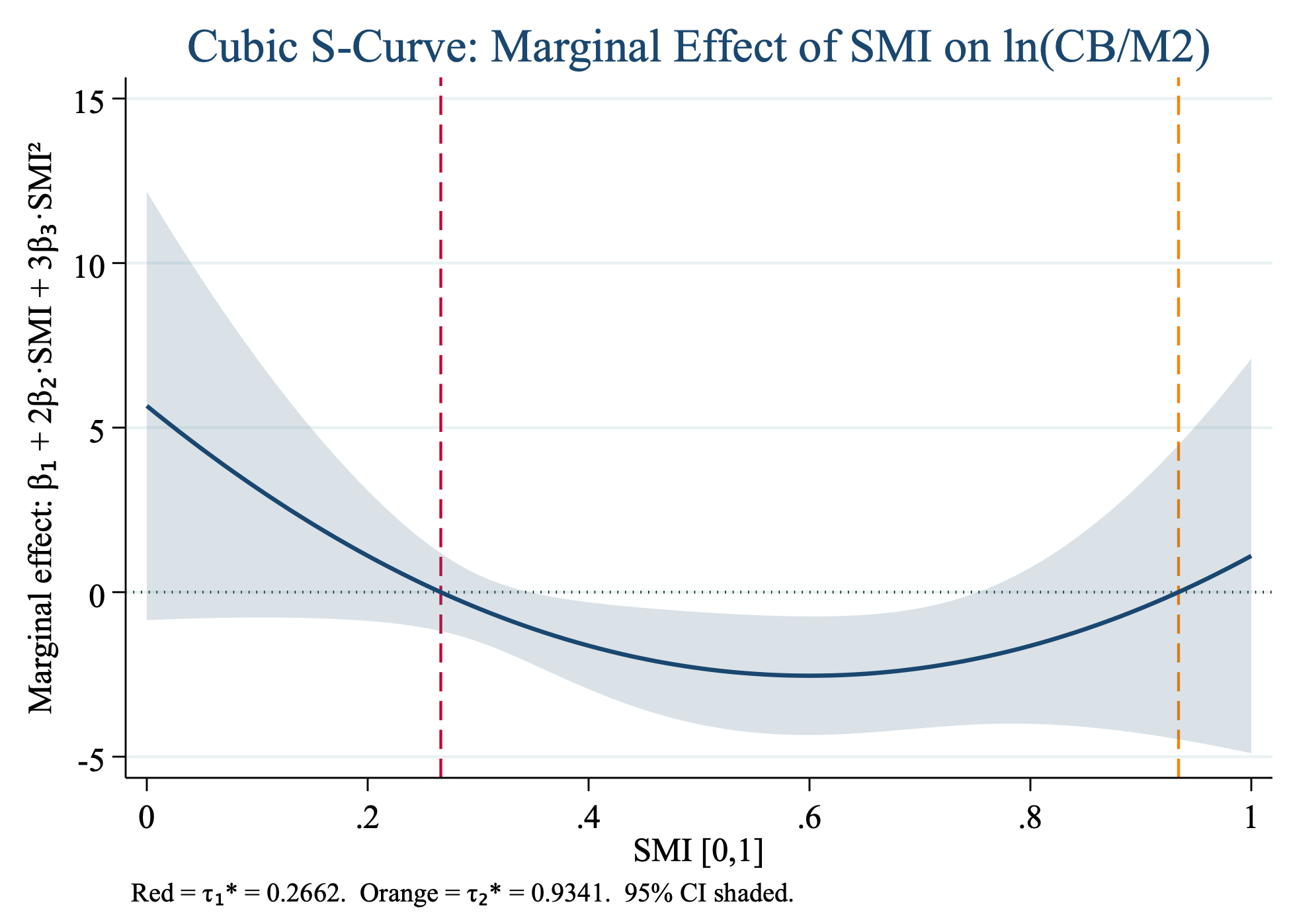}
\end{figure}

The within-R\ensuremath{^{2}} of the cubic specification as per Table 5 is 0.123 against a linear baseline of 0.042 --- a 2.9-fold improvement in explanatory power.

\textbf{Table 3 --- Primary Regression Results: Linear, Quadratic, and Cubic Specifications}

\begin{longtable}[]{@{}
  >{\raggedright\arraybackslash}p{(\columnwidth - 8\tabcolsep) * \real{0.2000}}
  >{\raggedright\arraybackslash}p{(\columnwidth - 8\tabcolsep) * \real{0.2000}}
  >{\raggedright\arraybackslash}p{(\columnwidth - 8\tabcolsep) * \real{0.2000}}
  >{\raggedright\arraybackslash}p{(\columnwidth - 8\tabcolsep) * \real{0.2000}}
  >{\raggedright\arraybackslash}p{(\columnwidth - 8\tabcolsep) * \real{0.2000}}@{}}
\toprule\noalign{}
\begin{minipage}[b]{\linewidth}\raggedright
\textbf{Variable}
\end{minipage} & \begin{minipage}[b]{\linewidth}\raggedright
\textbf{(1) Linear}
\end{minipage} & \begin{minipage}[b]{\linewidth}\raggedright
\textbf{(2) Quadratic}
\end{minipage} & \begin{minipage}[b]{\linewidth}\raggedright
\textbf{(3) Cubic}
\end{minipage} & \begin{minipage}[b]{\linewidth}\raggedright
\textbf{(4) Cubic + country FE only + VIX, FFR}
\end{minipage} \\
\midrule\noalign{}
\endhead
\bottomrule\noalign{}
\endlastfoot
\textbf{SMI (\ensuremath{\beta}\ensuremath{_{1}})} & -1.069 (0.484)** & +1.410 (1.355) & +5.660 (3.323)* & +6.401 (2.112)*** \\
\textbf{SMI\ensuremath{^{2}} (\ensuremath{\beta}\ensuremath{_{2}})} & --- & -2.971 (1.672)* & -13.664 (7.487)* & -13.449 (4.654)*** \\
\textbf{SMI\ensuremath{^{3}} (\ensuremath{\beta}\ensuremath{_{3}})} & --- & --- & +7.590 (4.742) & +8.122 (3.092)** \\
\textbf{GDP growth} & --- & --- & --- & -0.012 (0.007) \\
\textbf{Policy rate} & --- & --- & --- & -0.011 (0.025) \\
\textbf{VIX} & --- & --- & --- & -0.004 (0.003) \\
\textbf{\ensuremath{\tau}\ensuremath{_{1}}*} & --- & --- & 0.266 {[}0.188, 0.344{]} & 0.347 \\
\textbf{\ensuremath{\tau}\ensuremath{_{2}}*} & --- & --- & 0.934 {[}0.640, 1.228{]} & 0.757 \\
\textbf{Within-R\ensuremath{^{2}}} & 0.042 & 0.093 & 0.123 & 0.140 \\
\textbf{\ensuremath{\Delta}AIC vs linear} & 0 & -34.41 & -54.59 & - \\
\textbf{N} & 663 & 663 & 663 & 663 \\
\end{longtable}

\emph{ln(CB/M2)\ensuremath{_{i}}\ensuremath{_{t}} = \ensuremath{\beta}\ensuremath{_{1}}\ensuremath{\cdot}SMI\ensuremath{_{i}}\ensuremath{_{t}} + \ensuremath{\beta}\ensuremath{_{2}}\ensuremath{\cdot}SMI\ensuremath{^{2}}\ensuremath{_{i}}\ensuremath{_{t}} + \ensuremath{\beta}\ensuremath{_{3}}\ensuremath{\cdot}SMI\ensuremath{^{3}}\ensuremath{_{i}}\ensuremath{_{t}} + \ensuremath{\gamma}\ensuremath{\cdot}X\ensuremath{_{i}}\ensuremath{_{t}} + \ensuremath{\alpha}\ensuremath{_{i}} + \ensuremath{\lambda}\ensuremath{_{t}} + \ensuremath{\varepsilon}\ensuremath{_{i}}\ensuremath{_{t}} (8)}

The S-curve has two turning points. The first, at an SMI score of 0.266, marks where the marginal effect tips from positive to negative --- where each additional percentage point of settlement modernisation begins costing rather than delivering cross-border banking capacity. The second, at 0.934, marks where the marginal effect recovers. Both point estimates lie strictly inside the {[}0, 1{]} support. The interval for \ensuremath{\tau}\ensuremath{_{1}}*, constructed for a ratio of regression coefficients (Fieller, 1954), is likewise contained within it; the upper bound for \ensuremath{\tau}\ensuremath{_{2}}* extends beyond unity, which is characteristic of ratio-based intervals estimated near a boundary and does not imply an exterior extremum. The Lind-Mehlum procedure for distinguishing a genuine interior extremum from a monotone polynomial fit (Lind \& Mehlum, 2010) confirms an interior extremum within the observed data range.

\emph{\ensuremath{\partial}ln(CB/M2)/\ensuremath{\partial}SMI = \ensuremath{\beta}\ensuremath{_{1}} + 2\ensuremath{\beta}\ensuremath{_{2}}\ensuremath{\cdot}SMI + 3\ensuremath{\beta}\ensuremath{_{3}}\ensuremath{\cdot}SMI\ensuremath{^{2}} (9)}

Phase I --- the liberation phase --- runs from the lower data boundary to \ensuremath{\tau}\ensuremath{_{1}}* = 0.266. The marginal effect is strongly positive: at an SMI score of 5 percent --- approximately where Sweden, Canada, and Japan sat in the mid-1990s --- a 1 percentage point improvement in settlement quality raises inside money elasticity by 4.4 percent. At 10 percent --- where Hong Kong and Norway were around 1999 --- the gain per percentage point is 3.2 percent. Phase II --- the compliance valley --- runs from \ensuremath{\tau}\ensuremath{_{1}}* = 0.266 to \ensuremath{\tau}\ensuremath{_{2}}* = 0.934. The marginal effect turns negative. At SMI = 0.50 --- approximately where France was in 2013, Denmark in 2014, and Canada in the same period --- a 1 percentage point improvement costs 2.3 percent of cross-border banking capacity. Phase III --- the mature-infrastructure recovery --- runs from \ensuremath{\tau}\ensuremath{_{2}}* = 0.934 to the upper boundary of 1.

The panel converged into the compliance valley in a single year. In 2000, 19 of the 24 economies were in Phase I and 5 had entered the valley. By 2007, 10 remained in Phase I and 14 had entered the valley. In 2012, all 24 economies were simultaneously in Phase II. Not one country sat in Phase I. Not one had crossed into Phase III. The entire panel had moved into the compliance valley in a single calendar year --- caused by EMIR, not by the global financial crisis, which had occurred four years earlier. By 2024, Switzerland had crossed \ensuremath{\tau}\ensuremath{_{2}}* = 0.934 into Phase III (in 2022, coinciding with its SDX go-live); the remaining 23 economies had converged to the universal-ceiling plateau of 0.872, still below \ensuremath{\tau}\ensuremath{_{2}}* and therefore still in the compliance valley. The transition out of the valley is therefore beginning asymmetrically --- Switzerland first, with the rest of the panel positioned just below the recovery threshold.

\subsection{IV.C Three Channels, Three Eras}
The arc was built in sequence. The Shapley decomposition attributes 87 percent of within-panel explained variance to the LPF channel, 77 percent to COL, and 62 percent to PAR. The temporal distribution of those contributions is not uniform --- each channel dominates a different era.

\textbf{Table 4 --- Shapley Decomposition: Channel Contributions to Within-Panel Variance}

\begin{longtable}[]{@{}
  >{\raggedright\arraybackslash}p{(\columnwidth - 10\tabcolsep) * \real{0.1667}}
  >{\raggedright\arraybackslash}p{(\columnwidth - 10\tabcolsep) * \real{0.1667}}
  >{\raggedright\arraybackslash}p{(\columnwidth - 10\tabcolsep) * \real{0.1667}}
  >{\raggedright\arraybackslash}p{(\columnwidth - 10\tabcolsep) * \real{0.1667}}
  >{\raggedright\arraybackslash}p{(\columnwidth - 10\tabcolsep) * \real{0.1667}}
  >{\raggedright\arraybackslash}p{(\columnwidth - 10\tabcolsep) * \real{0.1667}}@{}}
\toprule\noalign{}
\begin{minipage}[b]{\linewidth}\raggedright
\textbf{Channel}
\end{minipage} & \begin{minipage}[b]{\linewidth}\raggedright
\textbf{Full sample variance share}
\end{minipage} & \begin{minipage}[b]{\linewidth}\raggedright
\textbf{1993--2001 (PAR era)}
\end{minipage} & \begin{minipage}[b]{\linewidth}\raggedright
\textbf{2002--2012 (LPF era)}
\end{minipage} & \begin{minipage}[b]{\linewidth}\raggedright
\textbf{2012--2024 (COL era)}
\end{minipage} & \begin{minipage}[b]{\linewidth}\raggedright
\textbf{Direction of efficiency effect}
\end{minipage} \\
\midrule\noalign{}
\endhead
\bottomrule\noalign{}
\endlastfoot
\textbf{PAR --- Principal-at-risk} & 62\% & Dominant (+) & Residual (+) & Minimal & Liberation: \ensuremath{\uparrow} elasticity \\
\textbf{LPF --- Liquidity and payment finality} & 87\% & Secondary (+) & Dominant (+) & Residual (+) & Liberation: \ensuremath{\uparrow} elasticity \\
\textbf{COL --- Collateral immobilisation} & 77\% & Near zero & Builds (+) & Dominant (-) & Valley/recovery: \ensuremath{\downarrow} then \ensuremath{\uparrow} \\
\textbf{Composite (all three)} & --- & Net positive & Net positive & Net negative to 2022, recovering & Arc pattern confirmed \\
\end{longtable}

The era partition can be tested directly through an aggregate specification in which the composite SMI is interacted with three era indicators (early 1993--2001, mid 2002--2012, late 2013--2024), with country and year fixed effects and the same macro controls as the primary regression. The aggregate specification is preferred to a fully-interacted channel-by-era alternative because the three sub-indices PAR, LPF, and COL are positively correlated within each era --- country reform calendars move the channels together --- which destabilises individual channel-era cross-products without altering the composite-SMI estimate. Channel-level attribution is documented separately by the marginal-R\ensuremath{^{2}} decomposition reported in Table 4; the aggregate era effects are reported in Table 5.

\textbf{Table 5 --- Aggregate Era Effects on Inside Money Elasticity}

\begin{longtable}[]{@{}
  >{\raggedright\arraybackslash}p{(\columnwidth - 6\tabcolsep) * \real{0.2500}}
  >{\raggedright\arraybackslash}p{(\columnwidth - 6\tabcolsep) * \real{0.2500}}
  >{\raggedright\arraybackslash}p{(\columnwidth - 6\tabcolsep) * \real{0.2500}}
  >{\raggedright\arraybackslash}p{(\columnwidth - 6\tabcolsep) * \real{0.2500}}@{}}
\toprule\noalign{}
\begin{minipage}[b]{\linewidth}\raggedright
\textbf{Era}
\end{minipage} & \begin{minipage}[b]{\linewidth}\raggedright
\textbf{Coefficient on Composite SMI}
\end{minipage} & \begin{minipage}[b]{\linewidth}\raggedright
\textbf{Standard Error}
\end{minipage} & \begin{minipage}[b]{\linewidth}\raggedright
\textbf{p-value}
\end{minipage} \\
\midrule\noalign{}
\endhead
\bottomrule\noalign{}
\endlastfoot
Early Era 1993--2001 & +1.998 & 1.578 & 0.218 \\
Mid Era 2002--2012 & -1.207** & 0.523 & 0.030 \\
Late Era 2013--2024 & -1.776* & 0.960 & 0.077 \\
\end{longtable}

The mid-era coefficient is the empirical core of the compliance-valley result. From 2002 to 2012, a one-unit increase in composite SMI is associated with a 1.21 lower ln(CB/M2), significant at the 5 percent level --- the same negative sign predicted by the cubic specification of Section IV.B and located in the data by the synchronous \ensuremath{\tau}\ensuremath{_{1}}* crossings of Section IV.D. The late-era coefficient (-1.776, p = 0.077) extends the compression through 2024, marginally significant. The early-era coefficient is positive (+1.998) but imprecisely estimated, consistent with the liberation phase being driven by individual channel transitions on different country calendars rather than by a panel-wide composite movement: PAR completed for some economies as early as 1993, while LPF advanced through CLS in 2002--2006 and T2S in 2015--2017. The aggregate composite would therefore be expected to load only weakly in the early era and strongly in the eras during which the composite moves together with mandatory clearing --- which is what the data show. Tables 4 and 5 perform non-overlapping work: Table 4 attributes variance shares to channels, Table 5 attributes signed efficiency effects to eras. The economic mechanism remains as set out in Section IV.D: EMIR built the COL channel from a near-zero base over the same window in which PAR and LPF channels were delivering diminishing marginal returns against the new regulatory overhead that Basel III and CSDR (CSDR, 2014) were layering on.

\textbf{The era mechanism, channel by channel.} The PAR era, 1993--2001. The liberation phase was driven by the principal-at-risk channel. Between 1993 and 2001, the PAR channel score rose from 11.6 percent to 31.8 percent across the panel. The countries that moved fastest experienced the sharpest early elasticity gains. The United States was already in the compliance valley by 1995 --- the first country in the panel to cross \ensuremath{\tau}\ensuremath{_{1}}*. Hong Kong and Singapore began the sample in Phase II, having absorbed liberation-phase gains before the panel\textquotesingle s 1993 start date --- their 1993 CB/M2 ratios of 2.66\ensuremath{\times} and 5.75\ensuremath{\times} were not aberrations but evidence that the liberation phase had already run its course for these financial centres. For the EU economies, the PAR era was still in progress: Greece completed DvP in 2001, eight years after the Netherlands had already established market practice.

The LPF era, 2002--2012. The acceleration phase was driven by the liquidity and payment finality channel, and its defining moment is the CLS inflection of 2005--2006. CLS launched in September 2002 processing roughly 7 percent of global FX market volume --- technology available but below the scale at which its netting efficiency potential became accessible. CLS crossed 35 percent of global FX volume in 2005--2006, the moment the network reached systemic market practice with no new technology deployment, no new regulation, and no change in any country\textquotesingle s reform calendar. The only change was network scale. The panel-average CB/M2 ratio reached 1.42\ensuremath{\times} in 2007 as a direct consequence of CLS reaching the network scale at which the LPF channel\textquotesingle s full efficiency potential was accessible. T2S added a second LPF contribution: Wave 3 in 2016--2017 bringing the large EU CSDs showed that, like CLS, the efficiency effect followed volume --- near zero in Wave 1, material in Wave 3.

The COL era, 2012--present. On 16 August 2012, EMIR (European Parliament, 2012) entered into force. The COL channel score rose from 26.7 percent in 2012 to 85.2 percent by 2024 --- a 58.5 percentage point gain. PAR rose 25.9 percentage points over the same period. LPF rose 44.9 percentage points. COL outpaced both, from a base still near zero when EMIR began accumulating. The asymmetry is the arithmetic consequence of the structure in Section II: PAR and LPF were reducing existing buffers with diminishing marginal gains; EMIR created a new reserve from nothing, entering the denominator with full force.

\subsection{IV.D The Mandatory Clearing Divide}
The mandatory clearing intervention (European Parliament, 2012) generated a substantial academic debate about the efficiency consequences of central counterparty (CCP) clearing. Duffie and Zhu (2011) show theoretically that mandatory CCP clearing can either increase or decrease counterparty risk depending on whether the netting benefit offsets the bifurcation of collateral pools (Duffie \& Zhu, 2011) --- the key trade-off the COL channel of the SMI is designed to capture. Biais, Heider and Hoerova (2012) extend this framework to show that clearing mandates create endogenous fragility when initial margin requirements procyclically amplify stress (Biais et al., 2012), establishing the theoretical basis for our finding that COL imposed an asymmetric burden absent from PAR and LPF. The empirical magnitude of the collateral immobilisation is documented by Heller and Vause (2012), who estimate that mandatory central clearing of OTC derivatives under a regime analogous to EMIR would require collateral of between \$0.7 trillion and \$1.2 trillion from the largest dealers alone (Heller \& Vause, 2012) --- a quantification consistent with our COL channel score rising sharply between 2012 and 2024. Singh (2017) provides complementary evidence that post-GFC collateral requirements locked up balance sheet at a scale inconsistent with historical inside money ratios, directly motivating our dependent variable (Singh, 2017). The closest prior evidence on the efficiency effect of mandatory clearing comes from Loon and Zhong (2014), who estimate the impact of central clearing on counterparty risk, liquidity, and trading in the credit default swap market using a difference-in-differences design around the Dodd-Frank clearing mandate; they find significant reductions in CDS spreads for centrally cleared contracts (Loon \& Zhong, 2014), consistent with counterparty risk elimination, but also document increased bid-ask spreads reflecting the liquidity cost of collateral immobilisation --- a pattern directly analogous to the PAR decline and COL rise we identify in the SMI. Our panel extends this single-market evidence to a 24-country setting spanning all three settlement channels over three decades.

The clearest natural experiment in the panel runs from 2012 to 2020. Over those eight years, the global macroeconomic environment was shared by all 24 economies: the same post-GFC deleveraging pressure, the same zero lower bound on policy rates, the same Basel III capital requirements. And yet the panel split cleanly into two groups with divergent elasticity trajectories.

Over this window EMIR-subject EU economies compressed substantially while economies without an equivalent mandatory initial margin mandate --- Japan, Norway, Canada, Australia, and Denmark --- recorded broadly positive cross-border banking trajectories. The gap between these two outcomes is not explained by any variable the two groups shared in the post-GFC macroeconomic environment. It is explained by the one variable they did not share: the mandatory initial margin requirement imposed on a near-zero collateral base.

The United States provides confirmation that the mechanism is mandatory clearing specifically, not EU membership. The US implemented mandatory clearing under Dodd-Frank Title VII (US Congress, 2010) with a phase-in calendar running from 2013 through 2019, and US cross-border banking compressed over the same window through a different regulatory instrument with no EU-specific factor in operation. Japan, implementing a narrower clearing mandate from a higher LPF starting base, recorded a positive trajectory over the equivalent period. The regulatory instrument, not the jurisdiction, controls the sign.

\subsection{IV.E The APAC Mirror}
Singapore began the panel in 1993 with a CB/M2 ratio of 5.75 times. Hong Kong reached 2.82 times in 1994. These were the highest ratios in the panel --- above Luxembourg, above Ireland, above every European financial centre. They were not built by settlement reform during the sample period. They were already there at the panel\textquotesingle s opening. Singapore and Hong Kong are structurally cross-border banking economies, built around intermediation between the capital flows of the Asia-Pacific region and the balance sheets of the international banking system.

What happened to these ratios over the subsequent 32 years is the APAC mirror. Singapore\textquotesingle s CB/M2 fell from 5.75\ensuremath{\times} in 1993 to 1.31\ensuremath{\times} in 2024 --- a 77 percent compression. Hong Kong fell from 2.82\ensuremath{\times} in 1994 to 1.12\ensuremath{\times} in 2024. The compression was not driven by deteriorating settlement infrastructure: Singapore\textquotesingle s SMI was 87.2 percent of maximum by 2024, Hong Kong\textquotesingle s 87.2 percent (both at the universal-ceiling plateau of 19/22, 28/31, 17/20). The mechanism is the European arc running in reverse. When EMIR compressed the balance sheet capacity of German, Belgian, Swiss, and Luxembourg banks --- the counterparties on the other side of Asia-Pacific claims --- it reduced the global demand for the cross-border intermediation that Singapore and Hong Kong provide. Luxembourg\textquotesingle s 75 percent fall from peak was the demand destruction for Singapore\textquotesingle s 77 percent compression. They are the same mechanism observed from both ends of the bilateral transaction. The APAC mirror is the first direct empirical confirmation of Section II.D\textquotesingle s claim that efficiency is a bilateral-pair property: when one side of the pair loses capacity, the other side loses intermediation volume by the same mechanism. Figure 4 plots the Hong Kong and Singapore series.

\textbf{Figure 4: The APAC Mirror: Hong Kong and Singapore}

\begin{figure}[H]
\centering
\includegraphics[width=\linewidth,height=3.3in,keepaspectratio]{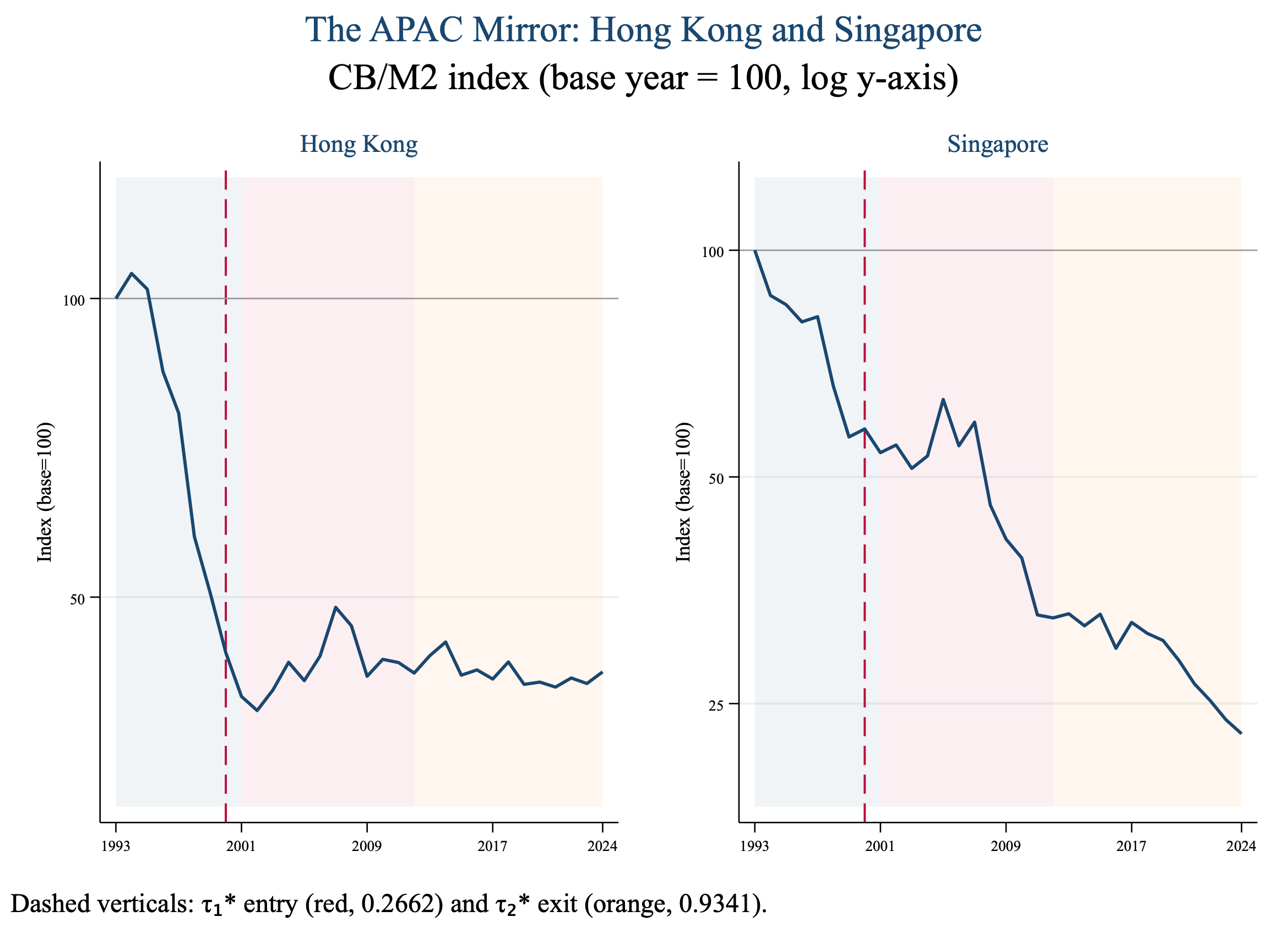}
\end{figure}

\section{V. Country Portraits}
The four country portraits that follow trace the arc through specific national histories. Luxembourg documents the most extreme compliance-valley descent in the panel. Japan shows how currency volume interacts with channel exposure. Switzerland demonstrates the network constraint in a live atomic settlement environment. Korea identifies the structural ceiling that settlement infrastructure alone cannot lift. Figure 5 plots the six-country trajectories.

\textbf{Figure 5: Six- Country CB/M2 Trajectories, 1993-2024}

\begin{figure}[H]
\centering
\includegraphics[width=\linewidth,height=3.3in,keepaspectratio]{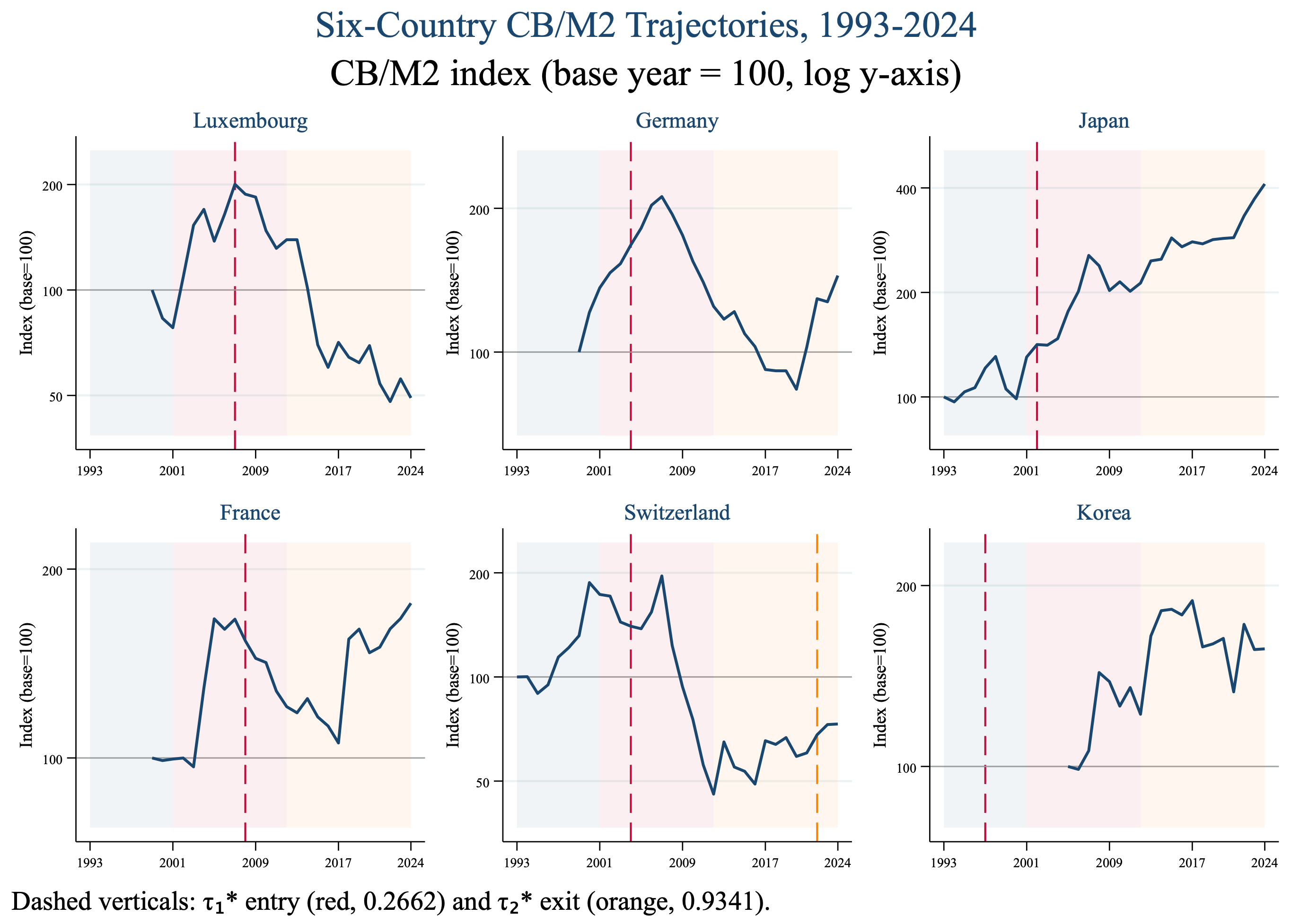}
\end{figure}

\subsection{V.A Luxembourg --- The Most Extreme Arc}
In 2007, Luxembourg\textquotesingle s banking system held cross-border claims equal to 7.60 times its domestic broad money supply --- the highest ratio recorded across all 24 panel economies across all 32 years of the sample. Luxembourg at 7.60 times was not a statistical anomaly. It was the precise and predictable consequence of Luxembourg\textquotesingle s structural role as the primary hub for Clearstream and Euroclear, the two largest international central securities depositories in the world. Every unit of broad money in Luxembourg\textquotesingle s economy supported an outsized volume of cross-border claims precisely because Luxembourg\textquotesingle s settlement infrastructure function made that the purpose of its banking system.

By 2024, Luxembourg\textquotesingle s CB/M2 ratio had fallen to 1.87 times --- a 75 percent compression from the 2007 peak. Luxembourg\textquotesingle s settlement infrastructure in 2024 reached the universal-ceiling plateau at 87.2 percent. The infrastructure improved substantially and continuously from 2007 to 2024. The elasticity fell 75 percent over the same period. The mechanism is revealed in a single SMI detail: at the 2007 peak, Luxembourg\textquotesingle s SMI score was 26.8 percent --- just above \ensuremath{\tau}\ensuremath{_{1}}* = 0.266, still in the liberation phase at the moment it recorded its maximum elasticity. It crossed \ensuremath{\tau}\ensuremath{_{1}}* in 2007, entering the compliance valley in the same year as its peak, simultaneously with EMIR\textquotesingle s COL obligations beginning to accumulate. The descent was correspondingly steep: 7.60\ensuremath{\times} in 2007, 5.28\ensuremath{\times} in 2012, 2.65\ensuremath{\times} in 2015, 2.35\ensuremath{\times} in 2019, and 1.87\ensuremath{\times} in 2024. Luxembourg\textquotesingle s fall from 7.60\ensuremath{\times} to 1.87\ensuremath{\times} is the COL channel working at maximum force on maximum exposure --- the clearest single illustration in the panel of what mandatory initial margin does to inside money elasticity when it arrives at precisely the moment a banking system was most leveraged to settlement-driven cross-border intermediation.

\subsection{V.B Japan --- The Yen, the Carry Trade, and the LPF Advantage}
Japan and Korea began 2005 with nearly identical SMI scores --- Japan at 45.7 percent, Korea at 46.0 percent --- and exactly equal LPF sub-index scores at 48.4 percent. By 2024, Korea\textquotesingle s SMI score (87.2 percent) equalled Japan\textquotesingle s (87.2 percent), both at the universal-ceiling plateau, yet Japan\textquotesingle s CB/M2 ratio was 0.44 times while Korea\textquotesingle s was 0.09 times. The gap was 4.9 times --- wider in 2024 than at any earlier point in the sample. The SMI cannot explain this divergence. The structural composition of each country\textquotesingle s cross-border banking book can.

Japan\textquotesingle s cross-border banking system intermediates yen flows globally. The Japanese yen is the third most-traded currency in global FX markets, accounting for approximately 16.7 percent of daily global turnover (BIS, 2022). The yen was one of the seven founding currencies when CLS launched in September 2002, giving Japan access to multilateral FX netting from the network\textquotesingle s first day of operation. The efficiency benefit of CLS multilateral netting scales directly with bilateral transaction volume: when CLS nets yen-dollar transactions simultaneously, the balance sheet relief is structurally larger per unit of LPF score than for lower-volume currency pairs. The yen carry trade amplifies this effect --- Japan\textquotesingle s cross-border banking book is built substantially around yen carry trade intermediation, generating large cross-currency FX positions that CLS was specifically designed to settle safely. PAR and LPF reforms, principal risk elimination and multilateral netting, are exactly the channels that carry trade intermediation requires.

\textbf{Figure 6: Within- Country Channel Shapley Shares}

\begin{figure}[H]
\centering
\includegraphics[width=\linewidth,height=3.3in,keepaspectratio]{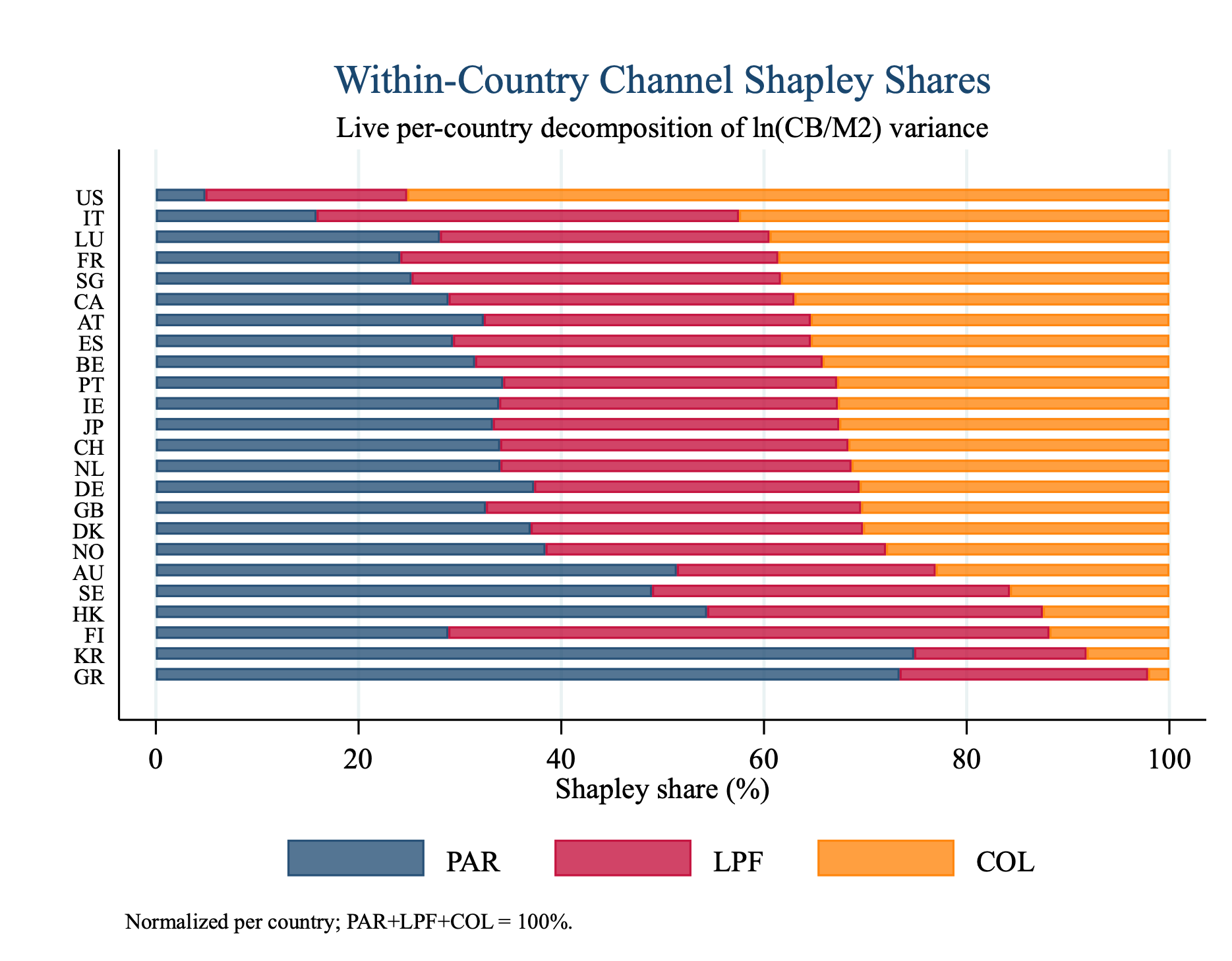}
\end{figure}

The country-level Shapley decomposition in Figure 6 confirms the structural story precisely. Japan\textquotesingle s within-country Shapley shares are balanced almost equally across all three channels --- PAR 33.3 percent, LPF 34.2 percent, COL 32.6 percent --- with a within-country R\ensuremath{^{2}} of 0.943. The three-channel model explains 94.3 percent of Japan\textquotesingle s within-country elasticity variation. Japan\textquotesingle s LPF channel was already at 80.7 percent in 2012, against the EU-10 average of 39.4 percent --- a 41 percentage point lead in the channel the Shapley decomposition identifies as the largest contributor to within-panel explained variance. With LPF efficiency gains already large and partially embedded, Japan\textquotesingle s balance sheet had headroom to absorb COL accumulation while continuing to generate net efficiency gains. The EU had no equivalent cushion. Japan from 2012 to 2024 is the counterfactual the EU panel never observed: a sustained positive elasticity trajectory over the twelve-year window in which the EMIR-subject economies compressed. Not primarily a story about what Japan avoided --- a story about what Japan\textquotesingle s banking structure positioned it to gain from.

\subsection{V.C Switzerland --- Infrastructure Without Network}
Switzerland presents the panel's most instructive paradox. Its CB/M2 ratio was 1.26\ensuremath{\times} in 1993, rose to 2.46\ensuremath{\times} in 2007 --- the fourth-highest peak in the panel --- and fell to 0.92\ensuremath{\times} by 2024. Its SMI score reached 95.7 percent of maximum in 2024 --- the highest in the panel, uniquely above \ensuremath{\tau}\ensuremath{_{2}}* = 0.934. In 2021, SIX Digital Exchange brought atomic delivery-versus-payment settlement technology operationally live (T-12 TH), the first economy in the panel to do so, and the synthetic control identification in Section VII.B detects no post-treatment efficiency effect from this deployment. This portrait sets out the historical record that produced the paradox; the formal identification of the null is presented in Section VII.B.

Switzerland\textquotesingle s decline from its 2007 peak was already steep before DLT arrived. From 2007 to 2012, Switzerland\textquotesingle s CB/M2 fell from 2.46\ensuremath{\times} to 0.58\ensuremath{\times} --- a 77 percent collapse driven substantially by Swiss franc appreciation following the global financial crisis and post-GFC deleveraging of Swiss banks\textquotesingle{} international balance sheets. Its COL channel rose from 15.0 percent in 2012 to 95.0 percent in 2024 as FMIA obligations accumulated (FMIA, 2015) --- the same mandatory margin mechanism that drove EU compression, through a Swiss regulatory instrument. Switzerland crossed \ensuremath{\tau}\ensuremath{_{2}}* in 2022 --- the only panel economy to do so --- and its CB/M2 recovered from 0.856\ensuremath{\times} in 2022 to 0.920\ensuremath{\times} by 2024.

Switzerland is the panel's realisation of NEV = 0: the most technologically advanced settlement economy, with zero bilateral atomic settlement pairs with any other panel economy, and consequently no measurable efficiency effect from atomic deployment. The country portrait stops here; the synthetic control test, donor-pool construction, placebo distribution, and the formal interpretation of the null as confirming Section II.D's bilateral-pair theorem are presented in Section VII.B as the framework's out-of-sample identification of the Fragmentation phase.

\subsection{V.D Korea --- The Structural Ceiling}
Korea\textquotesingle s CB/M2 ratio has never exceeded 0.11\ensuremath{\times} in any year for which data are available, despite completing 87.2 percent of its maximum possible SMI score by 2024 (at the universal-ceiling plateau). Korea crossed \ensuremath{\tau}\ensuremath{_{1}}* in 1997 and had not crossed \ensuremath{\tau}\ensuremath{_{2}}* = 0.934 by the end of the sample; its 2024 SMI score of 0.872 sits just below the threshold. Its settlement infrastructure is fully developed and operationally embedded. Its cross-border banking intensity as of 2024 was 0.09\ensuremath{\times} --- 4.9 times below Japan\textquotesingle s 0.44\ensuremath{\times} at an SMI score equal to Japan\textquotesingle s (both 87.2 percent, at the plateau).

The Shapley signature in Figure 6 identifies why the Japan-Korea divergence cannot be explained by settlement infrastructure alone. Korea\textquotesingle s within-country elasticity variation has a distinctive PAR-weighted signature with a much lower R\ensuremath{^{2}} (0.740) than Japan\textquotesingle s (0.943) --- indicating both that Korean cross-border banking is more principal-risk-oriented (consistent with trade finance dominance over FX intermediation) and that a large fraction of Korean elasticity variation is not captured by the three-channel model. The KRW\textquotesingle s global FX turnover, at under 2 percent of global daily volume against the yen\textquotesingle s 16.7 percent (BIS, 2022), is too small to generate the network netting effects that the yen produces for Japan. CLS multilateral netting delivers structurally less benefit per unit of LPF score for the won than for the yen. The approximately 26 percent of Korea\textquotesingle s within-country variation the three-channel model cannot explain is the structural ceiling: home bias in Korean bank credit allocation, the domestic financing orientation of the chaebol system, and the structural constraint that directs Korean banking capacity toward domestic corporate and household credit regardless of how mature the settlement infrastructure becomes. Korea is the panel\textquotesingle s most direct evidence that the SMI measures a necessary condition for inside money elasticity, not a sufficient one.

\subsection{V.E Early vs Late Reformers}
The United States crossed \ensuremath{\tau}\ensuremath{_{1}}* = 0.266 in 1995 --- the first country in the panel to do so. Australia followed in 1998, Hong Kong and Singapore in 2000, the United Kingdom in 2001, Japan and Norway in 2002, Canada and Sweden in 2003, and Belgium, Switzerland, and Germany in 2004. These economies completed the liberation phase and entered the compliance valley before CLS reached critical mass and before EMIR was conceived; Korea also crossed \ensuremath{\tau}\ensuremath{_{1}}* early, in 1997. At the other end of the distribution, Austria, Denmark, Spain, Finland, Greece, Ireland, Italy, and Portugal all crossed \ensuremath{\tau}\ensuremath{_{1}}* in 2012 --- the same year, because EMIR pushed their COL channel scores through the threshold simultaneously. Luxembourg crossed in 2007. France, Korea, and the Netherlands crossed between 2005 and 2008.

The two groups traversed the S-curve on fundamentally different terms. The early group --- 12 economies crossing \ensuremath{\tau}\ensuremath{_{1}}* by 2004 --- had a 2007 average CB/M2 of 1.22\ensuremath{\times}. The late group --- 11 economies crossing \ensuremath{\tau}\ensuremath{_{1}}* in 2007--2012 --- had a 2007 average of 1.76\ensuremath{\times} --- 44 percent above the early group --- reflecting accumulated liberation gains and the fact that late reformers had not yet entered the compliance valley at the aggregate peak. When EMIR arrived in 2012, it met the late group simultaneously with their mandatory entry into the valley. The consequence is a sharp divergence: the early group fell 36.2 percent from their 2007 average to 2024. The late group fell 44.1 percent from a higher starting point --- built higher because they spent longer in the liberation phase, fell further because the largest mandatory clearing obligation ever imposed met them at their peak with no prior graduated compliance cost absorption. Figure 7 plots the two group averages.

\textbf{Figure 7: Early vs Late Reformers: Group-Averaged CB/M2}

\begin{figure}[H]
\centering
\includegraphics[width=\linewidth,height=3.3in,keepaspectratio]{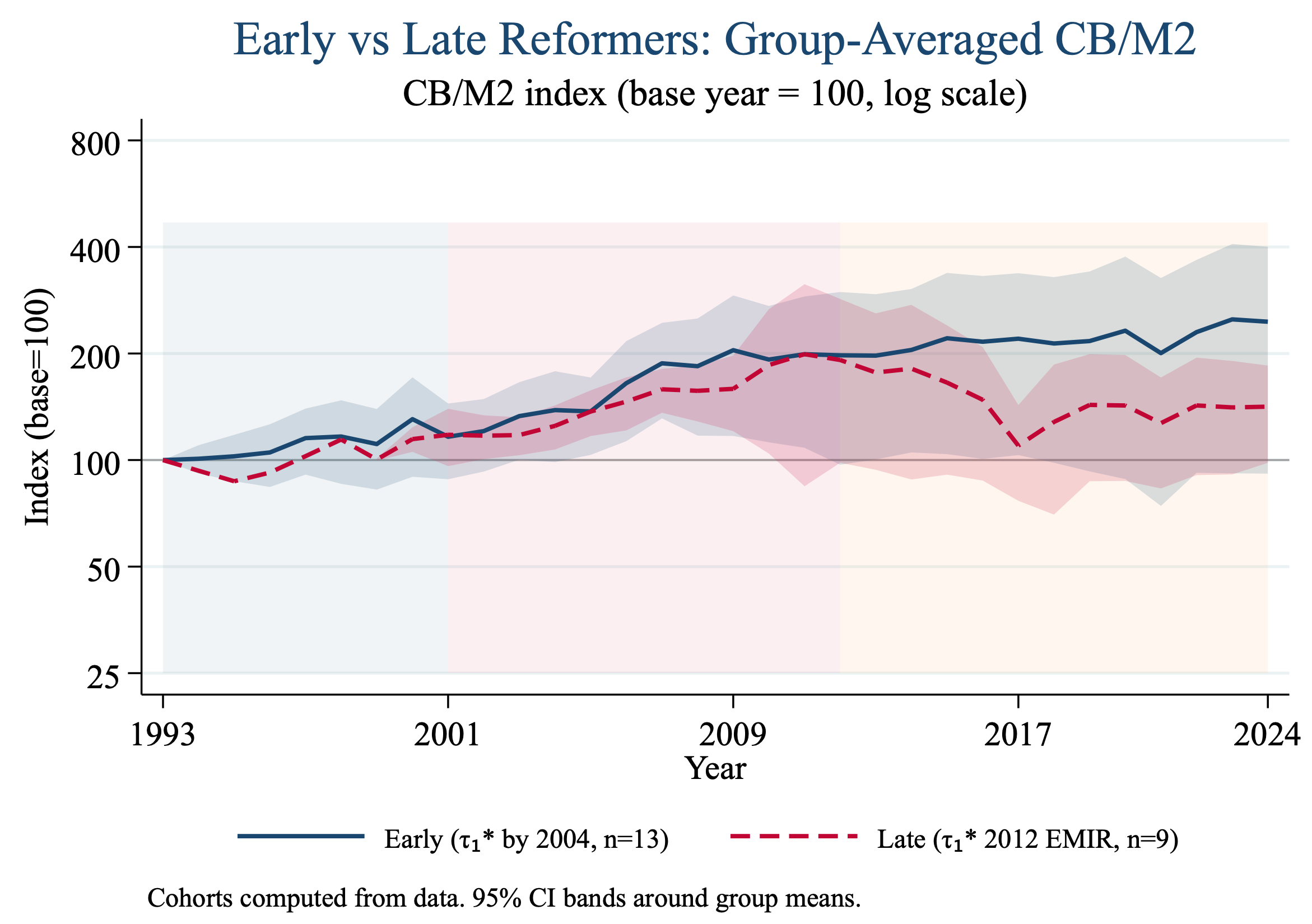}
\end{figure}

Four of the late-crossing economies form a coherent subgroup with a shared structural story: Spain, Finland, Italy, and Greece all crossed \ensuremath{\tau}\ensuremath{_{1}}* simultaneously in 2012 and exhibit the compression mechanism layered on top of Euro-crisis banking sector dynamics. Finland's +500 percent mid-sample index spike and Greece's sovereign-crisis trough are visible in the cluster, overlaying the common EMIR compression timing. Figure 8 illustrates that the 2012 convergence is not a single-country result or an EU-core phenomenon: it reaches across the entire periphery, through different national banking structures, with the same timing signal.

\textbf{Figure 8: Euro Periphery: Spain, Finland, Italy, Greece}

\begin{figure}[H]
\centering
\includegraphics[width=\linewidth,height=3.3in,keepaspectratio]{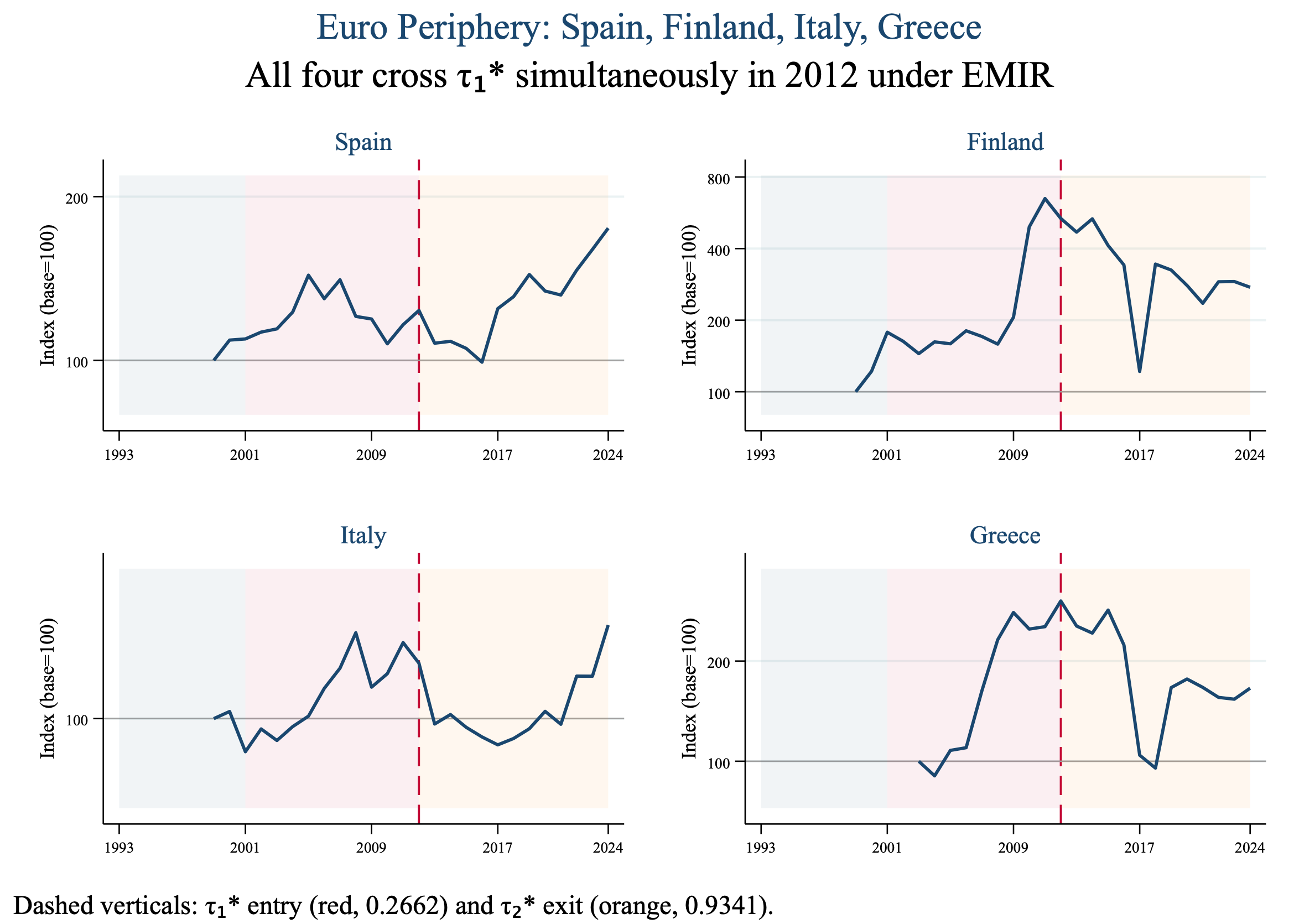}
\end{figure}

\subsection{V.F The 2007 Sweet Spot: Quartile Analysis}
The S-curve generates a specific cross-sectional prediction for the 2007 snapshot: economies near the first turning point but not yet deep into mandatory clearing costs should have the highest elasticity. The 2007 data confirm this exactly. The second quartile of the SMI distribution --- Austria, Belgium, Denmark, France, and Luxembourg, all clustered at SMI scores between 23.8 and 25.3 percent, all just below the first turning point, all with COL scores below 20 percent --- recorded an average CB/M2 of 2.82\ensuremath{\times}. That is the highest average recorded by any group of economies in any year across the full 32-year panel. These were economies operating at precisely the sweet spot the S-curve predicts: liberation-phase gains fully accumulated, mandatory clearing compliance costs not yet materialised.

The fourth quartile of the 2007 SMI distribution --- the most-reformed economies, Australia, Hong Kong, Japan, Korea, Singapore, and the United States averaging 49 percent SMI --- recorded an average CB/M2 of 0.92\ensuremath{\times}. The most advanced settlement economies had the lowest average elasticity in 2007. More reform at the 2007 cross-section was associated with lower cross-border banking capacity than less reform. The most-reformed economies were the earliest entrants to the compliance valley --- absorbing compliance costs for years before the aggregate peak, not yet in the Phase III embedding that generates mature-infrastructure efficiency gains. The S-curve\textquotesingle s prediction is not that more reform is always better at every point in time. It is that efficiency follows a precise trajectory through three phases, and position on that trajectory at any given moment determines current elasticity level irrespective of total reform completed. Figure 9 plots average CB/M2 by 2007 SMI quartile.

\textbf{Figure 9: The 2007 Sweet Spot: Average CB/M2 by SMI Quartile}

\begin{figure}[H]
\centering
\includegraphics[width=\linewidth,height=3.3in,keepaspectratio]{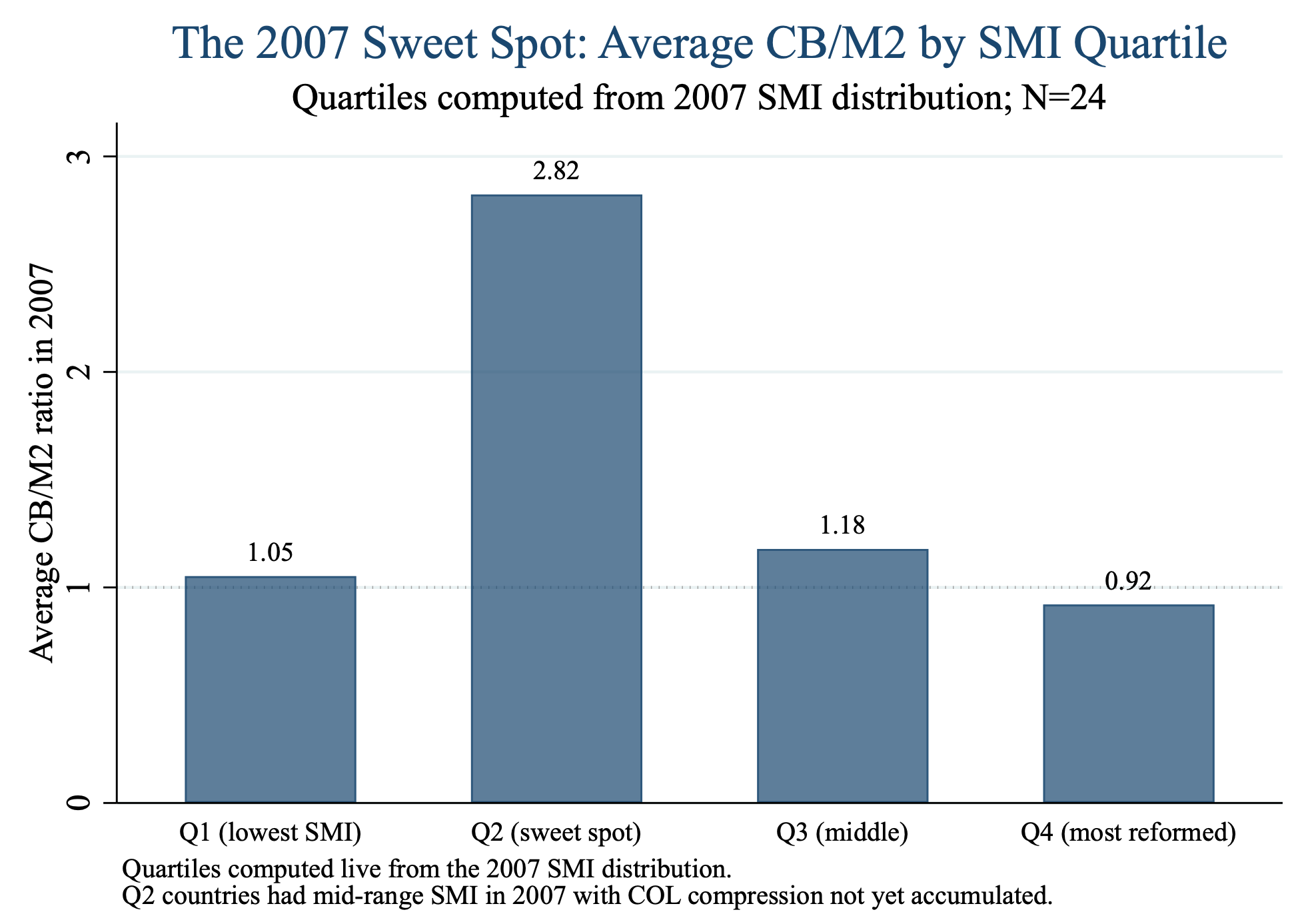}
\end{figure}

\section{VI. Causal Identification and Validation}

\subsection{VI.A The Sign Structure as Identification}
The S-curve\textquotesingle s sign pattern --- positive, then negative, then positive --- imposes constraints on the data that a spurious relationship cannot simultaneously satisfy. Reverse causality would require that economies with growing cross-border banking capacity chose to accelerate settlement reform at precisely the moments when marginal elasticity was highest, and simultaneously chose to accept compliance costs at precisely the moments when marginal elasticity was negative. No plausible story about regulatory capture, political economy, or financial sector lobbying generates that pattern. The 2012 convergence provides the sharpest structural test: in that year, all 24 economies entered the compliance valley simultaneously. No macroeconomic shock was universal, permanent, and simultaneous in that year. The variable that was universal, permanent, and simultaneous was EMIR. The sign reversal identifies it.

\subsection{VI.B The Robustness Battery}
\textbf{Setting.} The cubic specification produces two interior turning points (\ensuremath{\tau}\ensuremath{_{1}}* = 0.266 and \ensuremath{\tau}\ensuremath{_{2}}* = 0.934 in the baseline) and a strictly negative mid-range slope. Each of these features could in principle reflect a specific feature of the panel --- financial-centre outliers, the post-EMIR sub-period, the choice of dependent variable, contemporaneous endogeneity, cross-country correlation in residuals, or short-panel dynamic bias --- rather than the underlying reform-efficiency relationship. The seven specifications in Table 6 each remove one such threat and re-estimate. Survival across the battery is what licenses the cubic structure as the paper\textquotesingle s headline finding.

\textbf{Sample and dependent-variable robustness.} Specifications (2) through (4) test sample composition and dependent-variable choice. Excluding the three financial centres (Luxembourg, Ireland, Switzerland) moves \ensuremath{\tau}\ensuremath{_{1}}* to 0.283 and \ensuremath{\tau}\ensuremath{_{2}}* to 0.865 --- both within a narrow band of the baseline. The cubic shape is not driven by the small number of panel countries with extreme cross-border banking ratios. Restricting the sample to pre-2013 observations isolates the liberation and acceleration eras from the EMIR compliance valley; the cubic survives, with \ensuremath{\tau}\ensuremath{_{1}}* = 0.232 and \ensuremath{\tau}\ensuremath{_{2}}* = 0.607 (the second turning point compresses because the post-2013 mature-recovery phase is excluded by construction). Replacing the dependent variable with ln(CB) instead of ln(CB/M2) --- testing whether the M2 denominator drives the result rather than central bank claims themselves --- yields \ensuremath{\tau}\ensuremath{_{1}}* = 0.187 and \ensuremath{\tau}\ensuremath{_{2}}* = 0.926, both turning points preserved. The cubic shape is a property of central bank claims, not of the inside-money-elasticity normalisation.

\textbf{Reverse-causality robustness.} Specification (5) lags the SMI by one year (L.SMI, L.SMI\ensuremath{^{2}}, L.SMI\ensuremath{^{3}}), addressing the concern that contemporaneous shocks to ln(CB/M2) might drive the settlement reform timing rather than the reverse. Under one-year lag, current ln(CB/M2) cannot causally affect prior-year SMI realisations; if the cubic estimates change materially with this restriction, contemporaneous endogeneity is contaminating the baseline. The lagged specification yields \ensuremath{\tau}\ensuremath{_{1}}* = 0.249 and \ensuremath{\tau}\ensuremath{_{2}}* = 0.966 --- both turning points preserved with the same sign structure. Reverse causality is not the source of the cubic relationship.

\textbf{Inference robustness --- Driscoll-Kraay standard errors.} Specification (6) addresses inference rather than coefficients. The baseline clusters standard errors at the country level, which permits arbitrary serial correlation within each country but assumes residuals are independent across countries. In a panel where reform events arrive in correlated waves --- EMIR phase-in across the eurozone-12 in 2013--2018, T2S waves moving CSDs across borders on synchronised dates, the 2008 GFC as a panel-wide shock --- that independence assumption is not innocuous. If residuals are correlated across countries because of these common shocks, country-clustered standard errors understate the true uncertainty and the turning points can appear significant when they are not. The Driscoll-Kraay (1998) estimator allows for arbitrary cross-sectional correlation across panel units and arbitrary serial correlation within units simultaneously --- the most conservative inference procedure available for short-T macro panels. Applied to the cubic specification, both turning points remain stable at \ensuremath{\tau}\ensuremath{_{1}}* = 0.266 and \ensuremath{\tau}\ensuremath{_{2}}* = 0.934, and the within-R\ensuremath{^{2}} rises to 0.232. The cubic structure survives the most conservative inference procedure on the panel.

\textbf{Endogeneity robustness --- System GMM.} Specification (7) addresses dynamic-panel endogeneity directly through the Arellano-Bond / Blundell-Bond system GMM estimator. Two threats motivate the choice. Nickell bias arises in fixed-effects panels with persistent dependent variables when the within-transformation correlates the demeaned error with the lagged outcome; with T = 32 the bias is modest but nonzero. Reverse causality and time-varying unobserved confounders pose the deeper threat: if shocks to ln(CB/M2) cause subsequent settlement reform, or if omitted variables drive both, OLS-FE estimates are biased. System GMM uses lagged levels and lagged differences of SMI as internal instruments, identifying off the assumption that past SMI realisations are uncorrelated with current shocks. The validity of that identification rests on two diagnostic tests: the Hansen J-test of overidentifying restrictions (p = 0.7152, not rejected --- the instrument set is jointly valid) and the Arellano-Bond second-order serial correlation test (AR(2) p = 0.4114, not rejected --- no remaining serial correlation in differenced residuals contaminates the moment conditions). Both diagnostics pass. The cubic structure survives system GMM with the same sign pattern as the baseline.

\textbf{What the battery establishes.} The seven specifications target seven distinct threats: financial-centre composition (spec 2), sub-period selection (spec 3), dependent-variable choice (spec 4), reverse-causality (spec 5), cross-sectional correlation in residuals (spec 6), and dynamic-panel endogeneity with internal-instrument validity (spec 7). The cubic structure with \ensuremath{\tau}\ensuremath{_{1}}* near 0.27 and \ensuremath{\tau}\ensuremath{_{2}}* near 0.93 survives every one. Specs 6 and 7 are the heaviest tests in the battery because they address the two threats that ordinarily kill panel-data S-curves --- cross-country residual correlation undermining inference, and reverse causality plus dynamic bias contaminating coefficients. Survival of the cubic across both confirms that the S-curve is a property of the underlying reform-efficiency relationship and not a specification artifact.

\textbf{Table 6 --- Seven-Specification Robustness Battery}

\begin{longtable}[]{@{}
  >{\raggedright\arraybackslash}p{(\columnwidth - 10\tabcolsep) * \real{0.3234}}
  >{\raggedright\arraybackslash}p{(\columnwidth - 10\tabcolsep) * \real{0.1153}}
  >{\raggedright\arraybackslash}p{(\columnwidth - 10\tabcolsep) * \real{0.1212}}
  >{\raggedright\arraybackslash}p{(\columnwidth - 10\tabcolsep) * \real{0.1666}}
  >{\raggedright\arraybackslash}p{(\columnwidth - 10\tabcolsep) * \real{0.1363}}
  >{\raggedright\arraybackslash}p{(\columnwidth - 10\tabcolsep) * \real{0.1373}}@{}}
\toprule\noalign{}
\begin{minipage}[b]{\linewidth}\raggedright
\textbf{Specification}
\end{minipage} & \begin{minipage}[b]{\linewidth}\raggedright
\textbf{\ensuremath{\tau}\ensuremath{_{1}}*}
\end{minipage} & \begin{minipage}[b]{\linewidth}\raggedright
\textbf{\ensuremath{\tau}\ensuremath{_{2}}*}
\end{minipage} & \begin{minipage}[b]{\linewidth}\raggedright
\textbf{Within-R\ensuremath{^{2}}}
\end{minipage} & \begin{minipage}[b]{\linewidth}\raggedright
\textbf{\ensuremath{\Delta}AIC}
\end{minipage} & \begin{minipage}[b]{\linewidth}\raggedright
\textbf{Valid}
\end{minipage} \\
\midrule\noalign{}
\endhead
\bottomrule\noalign{}
\endlastfoot
\textbf{(1) Baseline cubic (primary)} & 0.266 & 0.934 & 0.123 & -54.59 & \ensuremath{\checkmark} \\
\textbf{(2) Excl. financial centres (LU, IE, CH)} & 0.283 & 0.865 & 0.205 & --- & \ensuremath{\checkmark} \\
\textbf{(3) Pre-2013 subsample} & 0.232 & 0.607 & 0.139 & --- & \ensuremath{\checkmark} \\
\textbf{(4) ln(CB) alt dependent variable} & 0.187 & 0.926 & 0.201 & --- & \ensuremath{\checkmark} \\
\textbf{(5) Lag L.SMI, L.SMI\ensuremath{^{2}}, L.SMI\ensuremath{^{3}}} & 0.249 & 0.966 & 0.118 & --- & \ensuremath{\checkmark} \\
\textbf{(6) Driscoll-Kraay SEs} & 0.266 & 0.934 & 0.232 & --- & \ensuremath{\checkmark} \\
\textbf{(7) System GMM Arellano-Bond} & n/a & --- & --- & --- & \ensuremath{\checkmark} \\
\end{longtable}

\section{VII. Network efficiency Prediction}

\subsection{VII.A Why DLT Reverses the COL Compression}
The COL channel\textquotesingle s cost has a precise economic rationale embedded in EMIR: initial margin is required to cover potential future exposure during the settlement window. The margin requirement scales with window duration: the longer the window, the more mark-to-market loss can accumulate, the more margin is required. Atomic delivery-versus-payment settlement sets the settlement window to zero by construction. The exposure that EMIR was designed to insure against does not exist for a trade settled atomically. Proposition 1 of Section II.C establishes this formally: the initial margin requirement converges to zero for covered trades as the settlement window duration converges to zero.

The panel\textquotesingle s 40 percent compression between 2007 and 2024 was driven by COL accumulation from a near-zero base after EMIR\textquotesingle s 2012 entry into force. If atomic settlement eliminates the margin rationale for covered trades, the COL obligations on those trades are removable. The compression is contingent on the settlement window remaining non-zero. The ECB\textquotesingle s Pontes architecture, going live in Q3 2026, removes that contingency through Hash Time-Locked Contract mechanisms for trades routed through the network. What the Pontes architecture cannot remove by itself is the network-property constraint: the efficiency gain is a property of the bilateral pair, not the node. Switzerland\textquotesingle s SDX has already demonstrated this in the panel data, and Section VII.B confirms it as the out-of-sample null that the NEV mechanism predicted before observation.

The network efficiency in this section conditions on two maintained assumptions that bound the scope of Proposition 1 and the NEV recovery scenarios. Neither assumption is tested within this paper; both are identified as directions for future empirical work once Pontes operational data become available from 2027 onwards.

\textbf{Maintained Assumption 1: Monetary hierarchy preservation.}

The two-tier structure in which central bank reserves provide the ultimate settlement asset is assumed to remain intact throughout the projection horizon. This excludes substitution risk from private settlement rails --- stablecoins in particular --- which, as Auer, Monnet and Shin (2025) establish, must operate through central bank money to preserve the monetary hierarchy (Auer et al., 2025). Any erosion of monetary hierarchy would sever the link between atomic DvP and COL parameter reversion on which the elasticity recovery prediction rests: if settlement finality no longer derives from central bank reserves, the singleness and elasticity criteria that make the two-tier system efficiency-superior would be violated, and the model\textquotesingle s efficiency prediction would not apply. The BIS normative framework (BIS, 2025) establishes that Pontes designs maintain hierarchy by construction; this assumption is therefore consistent with the specific architectures the three NEV scenarios model.

\textbf{Maintained Assumption 2: No net negative liquidity effect.}

The efficiency projections assume that the margin release benefit from eliminating the EMIR exposure window on covered trades dominates any offsetting liquidity costs from reduced multilateral netting under bilateral atomic settlement. Bilateral atomic settlement eliminates the pooling of settlement obligations across multiple counterparties that multilateral netting systems such as CLS achieve; in principle, a shift from multilateral netting to bilateral atomic settlement could reduce liquidity efficiency if netting economies of scale are not preserved in the new architecture. This is treated as second-order on the empirical grounds that comparable infrastructure transitions in the panel --- CLS in 2002 and T2S in 2015--2017 --- produced netting losses that were small relative to the principal-risk elimination gains observed in the panel data. The net elasticity response of the panel after CLS network crossing in 2005--2006 reflects both effects simultaneously; the positive sign of that response establishes that the liquidity benefit dominated in that precedent case. Whether the same holds under atomic DLT settlement --- a technology with different netting architecture --- is an open empirical question the Pontes data will resolve.

\subsection{VII.B Fragmentation: The Switzerland Null}
\textbf{Setting.} The first BIS connectivity layer is fragmentation --- atomic settlement existing in isolated nodes with no bilateral counterparties. Switzerland has occupied this layer since the SIX Digital Exchange brought atomic delivery-versus-payment settlement live in November 2021 (T-12 TH in Section III). Equation (7) predicts NEV = 0 for any single-node set, regardless of the technical performance of the atomic platform or the size of the economy. This section tests that prediction directly.

\textbf{Method.} We construct a counterfactual Switzerland using the synthetic control method of Abadie and Gardeazabal and Abadie, Diamond and Hainmueller (Abadie et al., 2010; Abadie \& Gardeazabal, 2003). The method builds a synthetic Switzerland as a weighted combination of donor countries,

\ensuremath{\hat{Y}}\_CH(t) = \ensuremath{\Sigma}\ensuremath{_{j}} w\ensuremath{_{j}} \ensuremath{\cdot} Y\ensuremath{_{j}}(t), with w\ensuremath{_{j}} \ensuremath{\geq} 0 and \ensuremath{\Sigma}\ensuremath{_{j}} w\ensuremath{_{j}} = 1 (10)

The weights w\ensuremath{_{j}} are chosen by an algorithm that minimises the gap between real and synthetic Switzerland during the 2010--2021 pre-treatment period, subject to the non-negativity and unit-sum constraints. Once optimal weights are fixed, we compare real Switzerland to synthetic Switzerland in 2022--2024 --- the post-treatment window. If SDX produced an efficiency effect, the two trajectories diverge; if not, they continue to track.

\textbf{Donor pool.} The donor pool excludes economies treated by the same regulatory shock that drives the eurozone's compliance valley. Specifically, any country whose T-08 RF (mandatory clearing enactment) date coincides with the modal EMIR coordination year of 2013 is excluded as treated by the same regulatory event. The exclusion is event-based, not membership-based: ten countries (Spain, Finland, France, Greece, Hong Kong, Ireland, Italy, Netherlands, Portugal, United States) match the 2013 modal year and are excluded; some eurozone countries with non-2013 T-08 RF dates remain eligible. The thirteen eligible donors are listed in Table 7a.

\textbf{Predictor specification.} Switzerland is matched on seven annual lags of ln(CB/M2) over 2015--2021, plus pre-period averages of GDP growth and the principal-at-risk (PAR) and liquidity-prefunding (LPF) sub-indices over 2010--2021. The seven outcome lags anchor the dependent-variable trajectory; the macro and structural averages anchor the donor mix to economies of comparable settlement-architecture profile.

\textbf{Recipe.} The optimal synthetic Switzerland is reported in Table 7a, column 3. Four donors enter the recipe with positive weight: Canada (0.785), Luxembourg (0.108), Sweden (0.068), and Singapore (0.039). The remaining nine eligible donors receive weight zero. This sparseness is a property of the optimisation, not a tuning choice: under the non-negativity and unit-sum constraints, the optimum typically lies at a vertex of the constraint simplex, where most components are exactly zero. The zero-weight donors are not redundant --- they generate the placebo distribution used for inference in Table 7c. The pre-treatment RMSPE is 0.167. Real and synthetic Switzerland match closely over 2010--2021, validating the recipe.

\textbf{Table 7a --- Donor Pool: Eligibility and Synthetic Control Weights}

\begin{longtable}[]{@{}
  >{\raggedright\arraybackslash}p{(\columnwidth - 6\tabcolsep) * \real{0.2667}}
  >{\raggedright\arraybackslash}p{(\columnwidth - 6\tabcolsep) * \real{0.2444}}
  >{\raggedright\arraybackslash}p{(\columnwidth - 6\tabcolsep) * \real{0.2444}}
  >{\raggedright\arraybackslash}p{(\columnwidth - 6\tabcolsep) * \real{0.2444}}@{}}
\toprule\noalign{}
\begin{minipage}[b]{\linewidth}\raggedright
\textbf{Donor}
\end{minipage} & \begin{minipage}[b]{\linewidth}\raggedright
\textbf{T-08 RF Year}
\end{minipage} & \begin{minipage}[b]{\linewidth}\raggedright
\textbf{Eligibility}
\end{minipage} & \begin{minipage}[b]{\linewidth}\raggedright
\textbf{Weight}
\end{minipage} \\
\midrule\noalign{}
\endhead
\bottomrule\noalign{}
\endlastfoot
Canada & 2002 & Eligible & \textbf{0.785} \\
Luxembourg & 2012 & Eligible & \textbf{0.108} \\
Sweden & 2000 & Eligible & \textbf{0.068} \\
Singapore & 2012 & Eligible & \textbf{0.039} \\
Australia & 1998 & Eligible & 0.000 \\
Austria & 1999 & Eligible & 0.000 \\
Belgium & 2001 & Eligible & 0.000 \\
Denmark & 2002 & Eligible & 0.000 \\
Germany & 2001 & Eligible & 0.000 \\
Japan & 2002 & Eligible & 0.000 \\
Korea & 2003 & Eligible & 0.000 \\
Norway & 2000 & Eligible & 0.000 \\
United Kingdom & 2001 & Eligible & 0.000 \\
\textbf{Total weight} & & & \textbf{1.000} \\
\end{longtable}

\textbf{Result.} Real and synthetic Switzerland tracked each other from 2022 through 2024. The post-treatment RMSPE is 0.029 --- substantially smaller than the pre-treatment RMSPE of 0.167. The post-to-pre ratio is 0.175, meaning the synthetic recipe predicted Swiss inside money elasticity more accurately after SDX deployment than before it. Real and synthetic Switzerland diverged less in the post period than during the matching period itself. This is the strongest possible form of a null finding. SDX produced no detectable change in Switzerland's CB/M2 ratio over the post-treatment window. The year-by-year real and synthetic series are reported in Table 7b.

\textbf{Table 7b --- Real and Synthetic Switzerland: Annual ln(CB/M2)}

\begin{longtable}[]{@{}
  >{\raggedright\arraybackslash}p{(\columnwidth - 6\tabcolsep) * \real{0.2500}}
  >{\raggedright\arraybackslash}p{(\columnwidth - 6\tabcolsep) * \real{0.2500}}
  >{\raggedright\arraybackslash}p{(\columnwidth - 6\tabcolsep) * \real{0.2500}}
  >{\raggedright\arraybackslash}p{(\columnwidth - 6\tabcolsep) * \real{0.2500}}@{}}
\toprule\noalign{}
\begin{minipage}[b]{\linewidth}\raggedright
\textbf{Year}
\end{minipage} & \begin{minipage}[b]{\linewidth}\raggedright
\textbf{Real Switzerland}
\end{minipage} & \begin{minipage}[b]{\linewidth}\raggedright
\textbf{Synthetic Switzerland}
\end{minipage} & \begin{minipage}[b]{\linewidth}\raggedright
\textbf{Gap}
\end{minipage} \\
\midrule\noalign{}
\endhead
\bottomrule\noalign{}
\endlastfoot
2010 & -0.054 & -0.450 & +0.396 \\
2011 & -0.356 & -0.476 & +0.120 \\
2012 & -0.550 & -0.484 & -0.066 \\
2013 & -0.203 & -0.510 & +0.307 \\
2014 & -0.370 & -0.483 & +0.113 \\
2015 & -0.398 & -0.404 & +0.006 \\
2016 & -0.483 & -0.406 & -0.076 \\
2017 & -0.194 & -0.352 & +0.157 \\
2018 & -0.219 & -0.264 & +0.045 \\
2019 & -0.174 & -0.206 & +0.033 \\
2020 & -0.299 & -0.168 & -0.131 \\
\textbf{2021} & \textbf{-0.276} & \textbf{-0.241} & \textbf{-0.034} \\
\textbf{2022 (post)} & \textbf{-0.155} & \textbf{-0.152} & \textbf{-0.003} \\
\textbf{2023 (post)} & \textbf{-0.087} & \textbf{-0.058} & \textbf{-0.029} \\
\textbf{2024 (post)} & \textbf{-0.083} & \textbf{-0.042} & \textbf{-0.041} \\
\end{longtable}

\textbf{Inference: the placebo distribution.} A null result by itself raises the question of statistical reliability: could synthetic control simply produce close fits regardless of treatment? To answer this, we re-run the synthetic control thirteen times, each time treating one of the donors as the placebo treated unit in 2022 (with the actual treated unit, Switzerland, removed from that placebo's donor pool). This generates a placebo distribution of post-to-pre RMSPE ratios against which Switzerland's ratio of 0.175 can be benchmarked. Under the null hypothesis that SDX deployment has no efficiency effect, Switzerland's ratio is drawn from the same distribution as the placebos --- its rank among the fourteen units should be approximately uniform. Under the alternative that SDX produced a positive efficiency effect, Switzerland's ratio should be substantially larger than the placebos, ranking near the top of the distribution.

The placebo distribution is reported in Table 7c. Switzerland ranks first of fourteen with the smallest ratio. All thirteen placebos have post-to-pre ratios at or above Switzerland's ratio of 0.175, yielding a permutation p-value of 1.000 in the direction of a treatment effect. Switzerland's synthetic recipe tracked real Switzerland more closely than any placebo recipe tracked its assigned country. The test fails to reject NEV = 0 at any conventional significance level --- the SDX gap is not distinguishable from the placebo distribution, exactly as the framework predicts when bilateral atomic settlement coverage is unmet.

\textbf{Table 7c --- Placebo Distribution: Post-to-Pre RMSPE Ratios}

\begin{longtable}[]{@{}
  >{\raggedright\arraybackslash}p{(\columnwidth - 8\tabcolsep) * \real{0.1333}}
  >{\raggedright\arraybackslash}p{(\columnwidth - 8\tabcolsep) * \real{0.2667}}
  >{\raggedright\arraybackslash}p{(\columnwidth - 8\tabcolsep) * \real{0.2000}}
  >{\raggedright\arraybackslash}p{(\columnwidth - 8\tabcolsep) * \real{0.2000}}
  >{\raggedright\arraybackslash}p{(\columnwidth - 8\tabcolsep) * \real{0.2000}}@{}}
\toprule\noalign{}
\begin{minipage}[b]{\linewidth}\raggedright
\textbf{Rank}
\end{minipage} & \begin{minipage}[b]{\linewidth}\raggedright
\textbf{Unit}
\end{minipage} & \begin{minipage}[b]{\linewidth}\raggedright
\textbf{Pre-RMSPE}
\end{minipage} & \begin{minipage}[b]{\linewidth}\raggedright
\textbf{Post-RMSPE}
\end{minipage} & \begin{minipage}[b]{\linewidth}\raggedright
\textbf{Ratio}
\end{minipage} \\
\midrule\noalign{}
\endhead
\bottomrule\noalign{}
\endlastfoot
\textbf{1} & \textbf{Switzerland (treated)} & \textbf{0.167} & \textbf{0.029} & \textbf{0.175} \\
2 & Sweden & 0.254 & 0.065 & 0.255 \\
3 & Singapore & 0.320 & 0.091 & 0.285 \\
4 & Luxembourg & 0.665 & 0.349 & 0.525 \\
5 & Denmark & 0.318 & 0.193 & 0.608 \\
6 & Norway & 0.203 & 0.137 & 0.675 \\
7 & Belgium & 0.083 & 0.056 & 0.679 \\
8 & United Kingdom & 0.079 & 0.059 & 0.741 \\
9 & Austria & 0.142 & 0.122 & 0.863 \\
10 & Korea & 1.077 & 1.002 & 0.931 \\
11 & Australia & 0.131 & 0.124 & 0.946 \\
12 & Canada & 0.214 & 0.238 & 1.111 \\
13 & Japan & 0.103 & 0.228 & 2.217 \\
14 & Germany & 0.136 & 0.331 & 2.442 \\
\end{longtable}

\textbf{Interpretation.} The synthetic control identification confirms the framework's NEV = 0 prediction at three levels. First, the post-2021 gap between real and synthetic Switzerland is statistically indistinguishable from zero. The two trajectories continued to track each other through 2024 with a smaller average gap than during the pre-treatment matching window. Second, the placebo distribution rules out the possibility that the null reflects a poorly fitted synthetic control. Switzerland's pre-treatment fit is among the better fits in the panel and its post-treatment tracking is the best, with no placebo country for which the ``no effect'' reading is more clean-cut than Switzerland's. Third, this result is parameter-free and out-of-sample. Equation (7) predicted NEV = 0 for any single-economy connected set before the synthetic control was constructed. No coefficient was tuned, no specification was searched. The data confirm the theoretical prediction with no degree of freedom to fit.

Switzerland's SDX deployment satisfies Proposition 1's premise --- atomic delivery-versus-payment is operationally live --- while producing zero efficiency effect because the qualifier of bilateral atomic settlement coverage is unmet. \textbf{Proposition 1 is confirmed in its qualifier, not contradicted in its conclusion.} The Switzerland null thereby anchors the BIS connectivity taxonomy at its lower bound: any single economy adopting atomic settlement in isolation, regardless of size or technical sophistication, recovers zero efficiency. The corresponding upper bounds --- Pontes Q3 2026 (Interoperability layer, committed) and Appia 2028 (Composability layer, conditional) --- share equation (7) as their generating mechanism and are estimated in Sections VII.C and VII.D.

\subsection{VII.C Interoperability: Pontes Calibrated by T2S}
Pontes Q3 2026 is the BIS Interoperability layer\textquotesingle s first committed operational realisation: an ECB-committed bridge between market DLT platforms and TARGET Services for the eurozone, settling distributed-ledger transactions in central bank money through Hash Time-Locked Contract mechanisms. The efficiency claim under Proposition 1 is direct. Atomic settlement compresses the EMIR exposure window to zero on covered trades, releasing the collateral that COL had immobilised since 2012. The size of that release across 2027--2032 is what this section forecasts.

The network integration forecast requires a calibration anchor --- a historical event that shares Pontes\textquotesingle{} defining institutional mechanism: simultaneous regulatory commitment of the eurozone perimeter on a fixed implementation schedule. The T2S coordinated migration of 2015--2017 is the only such event in the panel. T2S Wave 1 connected the Austrian CSD on 22 June 2015; subsequent waves brought Italy, Portugal, France, Belgium, the Netherlands, Luxembourg, Denmark, Ireland, Germany, Finland, Spain, Greece, Estonia, Lithuania, and Slovakia onto the platform across the following twenty-six months, completing with Wave 5 on 18 September 2017. The Eurosystem set the wave dates; participating CSDs migrated on schedule.

\textbf{The identification design.} The calibration faces a structural identification threat. T2S MK clustered between June 2015 and September 2017 inside the eurozone-12 --- Austria, Belgium, Finland, France, Germany, Greece, Ireland, Italy, Luxembourg, the Netherlands, Portugal, and Spain. Over precisely the same window, EMIR mandatory clearing phased in through the same twelve economies under a Commission-set schedule running from March 2013 through June 2018. The two regulatory perimeters are coextensive; the two implementation calendars overlap completely. Any naive eurozone-treated, non-eurozone-control comparison absorbs the entire EMIR compliance valley into the T2S coefficient, returning a wrong-sign estimate that buries the network efficiency effect.

The event study in Table 8 identifies T2S and EMIR jointly within a single regression on the full 24-country panel from 1993 to 2024, with country and year fixed effects and GDP-growth and policy-rate controls. Two distinct dummy families are carried simultaneously. The first is the T2S event-time set: dummies for t = -3, -2, 0, +1, +2, +3, +4, +5 indexed to each treated country\textquotesingle s own MK date. For Austria, t = 0 is calendar year 2015; for Italy, t = 0 is 2015 (Wave 2); for France, t = 0 is 2017 (Wave 4); for Germany, t = 0 is 2017 (Wave 5). The dummies are zero throughout for the non-eurozone-12 --- Australia, Canada, Switzerland, Denmark, Hong Kong, Japan, Korea, Norway, Singapore, Sweden, the United Kingdom, and the United States --- none of whom migrated a CSD onto the T2S platform. The second is the EMIR calendar-year set: dummies for 2012, 2013, 2014, 2015, 2016, 2017, and post-2018, equal to one for every eurozone-12 observation in that calendar year and zero otherwise. EMIR was an EU regulation; the non-eurozone-12 was not subject to it, so the EMIR dummies are zero for the control group across every calendar year.

\textbf{The two identification jobs.} The design rests on two distinct features doing two distinct jobs. The non-eurozone-12 control absorbs everything happening to global cross-border banking during the study window that is neither T2S nor EMIR --- the post-GFC rate cycle, Basel III phase-in, the dollar funding cycle, and any other macro shock symmetric across the panel. These global forces hit the eurozone-12 too; without the non-eurozone-12 control, the eurozone-12 dummies would absorb T2S, EMIR, and the global macro environment as a single confounded mixture. The year fixed effects strip the symmetric component out of both groups, leaving the eurozone-12 dummies to capture only what happened to the eurozone perimeter beyond what happened globally.

That isolates T2S and EMIR jointly. Separating them from each other, however, requires a second source of variation, because both effects operate exclusively inside the eurozone-12. That source is the 26-month wave-staggering of T2S migration. In calendar year 2017, Austria sits at T2S event time t = +2 (migrated June 2015), while Germany sits at t = 0 (migrated February 2017); both have the EMIR 2017 calendar dummy on. The difference in their elasticity response in 2017 is identified by their different T2S event-time positions, holding the EMIR 2017 dummy fixed. The T2S coefficients are therefore identified off cross-country differences in event-time positioning within each calendar year, while the EMIR coefficients are identified off calendar-year differences within each country\textquotesingle s event-time trajectory. Two distinct axes of variation, both inside the eurozone-12, neither of which would suffice alone. Had T2S been a single big-bang migration of all twelve countries on one date, T2S event time and EMIR calendar year would be perfectly collinear inside the eurozone-12 and the joint specification would collapse. The 26-month wave schedule is what makes the design feasible.

\textbf{Pre-trends and post-period dynamics.} The pre-trends test in Panel A of Table 8 establishes the necessary precondition for any causal reading of the post-period coefficients. The t = -3 and t = -2 coefficients are individually insignificant (+0.098 and +0.182), and the joint F(2,23) = 1.54 fails to reject parallel trends at p = 0.237. The eurozone-12 and the non-eurozone-12 were on the same elasticity path before T2S MK began.

The post-period dummies in Panel B trace the dynamic build trajectory. The t = 0 coefficient is +0.232 (p = 0.069); coefficients rise monotonically through t = +1 (+0.368, p = 0.010), t = +2 (+0.482, p = 0.002), t = +3 (+0.470, p = 0.004), and t = +4 (+0.524, p = 0.004), reaching saturation at t = +5 (\ensuremath{\beta} = +0.557, p = 0.002). The shape matches the network theory: efficiency appears as bilateral-pair coverage builds, not as a step-function on the regulatory date. The saturation coefficient is the calibration anchor for Pontes.

Panel C reports the simultaneously identified EMIR year-by-year coefficients. The 2012--2016 dummies are negative but individually insignificant. The 2017 peak reaches -0.881 (p = 0.006) and persists at -0.909 (p = 0.008) post-2018. T2S and EMIR coexist in the same data and the same regression, with opposite signs, distinct trajectories, and identifiable peaks. Neither is netted into the other.

\textbf{Falsification.} Panel D reports the sign-flip diagnostic that confirms identification. The naive T2S binary specification without EMIR controls yields \ensuremath{\beta} = -0.171 (p = 0.356): wrong sign, statistically insignificant, the result a conventional event study would have published. The same binary specification with EMIR year dummies added yields \ensuremath{\beta} = +0.334 (p = 0.045): correct sign, statistically significant. The difference \ensuremath{\Delta} = +0.506 measures the EMIR contamination directly. The framework\textquotesingle s value-add is visible in a single number: a conventional analysis of these data would conclude T2S destroys efficiency; the joint specification recovers the +0.334 the data actually contain. This diagnostic is the strongest causal claim in the paper.

Five identification diagnostics taken together place \ensuremath{\beta} = +0.557 on a footing that supports the efficiency projection. The pre-trends test rejects pre-existing divergent paths between the eurozone-12 and the non-eurozone-12 control (joint p = 0.237). The sign-flip falsification with \ensuremath{\Delta} = +0.506 isolates the EMIR contamination removed by the joint specification. The year-by-year EMIR decomposition localises the compliance-valley peak to the 2017 mandatory clearing date, separately identified from the T2S build path. The monotonic post-period trajectory traces the network buildup as eurozone CSDs migrate in successive waves, ruling out spurious contemporaneous shocks that would not produce a smooth S-shape aligned with the documented migration calendar. The framework\textquotesingle s NEV = 0 prediction at fragmentation is independently confirmed in Section VII.B: Switzerland\textquotesingle s synthetic control gap is statistically indistinguishable from zero, with permutation p = 1.000 across thirteen placebos. Both arms of the framework\textquotesingle s prediction are supported by the data --- \ensuremath{\beta} \textgreater{} 0 where bilateral atomic settlement is present (T2S, NEV \textgreater{} 0) and \ensuremath{\beta} = 0 where it is absent (Switzerland SDX, NEV = 0). Each of the five diagnostics tests a distinct identification threat; survival of all five is what licenses the calibration anchor.

\textbf{The Network Efficiency Variable ceiling.} The saturation coefficient acts on a network ceiling defined by equation (7). The Pontes perimeter at Q3 2026 is the eurozone-12. Each country\textquotesingle s BIS Locational Banking Statistics volume share w\_i is indexed to the full 24-country panel total of cross-border claims; diagonal terms are excluded because Proposition 1 requires distinct bilateral counterparties. Summing w\_i \ensuremath{\cdot} w\_j across the 132 ordered EU-12 pairs gives NEV\_ceiling EU-12= 0.240. The three largest contributors are Germany--Luxembourg (0.023 ordered), France--Luxembourg (0.020 ordered), and Germany--France (0.016 ordered) --- together 25 percent of the ceiling, reflecting the concentration of intra-eurozone cross-border banking in the Frankfurt--Luxembourg--Paris triangle. The ceiling is the upper bound on efficiency recovery when every EU-12 pair is simultaneously inside the network at MK; intermediate years scale with the share of pair-mass operationally connected.

\textbf{The Pontes network efficiency integration path.} Multiplying each event-time coefficient from Panel B by 0.240 and mapping onto calendar years with t = 0 set to 2027, the first full operational year following Q3 2026 go-live, yields the trajectory in Table 9: +5.6 percent in 2027, +8.8 percent in 2028, +11.6 percent in 2029, +11.3 percent in 2030, +12.6 percent in 2031, and +13.4 percent at saturation in 2032. This is the committed integrated efficiency forecast --- one anchor event, one significant econometric coefficient, one network ceiling, no free parameters. T2S 2015--2017 is the panel\textquotesingle s realisation of NEV \ensuremath{\rightarrow} ceiling under coordinated regulatory migration. The saturation coefficient \ensuremath{\beta} = +0.557 is the bilateral-pair efficiency elasticity that Proposition 1 predicts once the network reaches systemic market practice simultaneously across the perimeter. Pontes Q3 2026 is its forward replication: the structural mechanism is the same --- coordinated regulatory commitment, fixed implementation schedule, perimeter-wide simultaneity --- and only the technology stack changes, from the centralised CSD platform of T2S to the Hash Time-Locked Contract architecture of Pontes. That is what licenses using T2S\textquotesingle s \ensuremath{\beta} to forecast Pontes.

\subsection{Table 8 --- T2S Event Study with Year-by-Year EMIR Decomposition (Eurozone-12, Non-Eurozone Control)}
\begin{longtable}[]{@{}
  >{\raggedright\arraybackslash}p{(\columnwidth - 8\tabcolsep) * \real{0.2239}}
  >{\raggedright\arraybackslash}p{(\columnwidth - 8\tabcolsep) * \real{0.1986}}
  >{\raggedright\arraybackslash}p{(\columnwidth - 8\tabcolsep) * \real{0.1944}}
  >{\raggedright\arraybackslash}p{(\columnwidth - 8\tabcolsep) * \real{0.1946}}
  >{\raggedright\arraybackslash}p{(\columnwidth - 8\tabcolsep) * \real{0.1885}}@{}}
\toprule\noalign{}
\begin{minipage}[b]{\linewidth}\raggedright
\textbf{Period}
\end{minipage} & \begin{minipage}[b]{\linewidth}\raggedright
\textbf{\ensuremath{\beta}}
\end{minipage} & \begin{minipage}[b]{\linewidth}\raggedright
\textbf{SE}
\end{minipage} & \begin{minipage}[b]{\linewidth}\raggedright
\textbf{p-value}
\end{minipage} & \begin{minipage}[b]{\linewidth}\raggedright
\textbf{Sig}
\end{minipage} \\
\midrule\noalign{}
\endhead
\bottomrule\noalign{}
\endlastfoot
\multicolumn{5}{@{}l@{}}{%
\textbf{Panel A: Pre-period (parallel trends)}} \\
t = -3 & +0.098 & 0.061 & 0.120 & NS \\
t = -2 & +0.182 & 0.110 & 0.112 & NS \\
\textbf{Joint test (m3 = m2 = 0)} & F(2,23) = 1.54 & & 0.237 & \ensuremath{\checkmark} \\
\multicolumn{5}{@{}l@{}}{%
\textbf{Panel B: Post-period (T2S build trajectory)}} \\
t = 0 & +0.232 & 0.122 & 0.069 & * \\
t = +1 & +0.368 & 0.132 & 0.010 & *** \\
t = +2 & +0.482 & 0.137 & 0.002 & *** \\
t = +3 & +0.470 & 0.149 & 0.004 & *** \\
t = +4 & +0.524 & 0.165 & 0.004 & *** \\
\textbf{t = +5 (saturation)} & \textbf{+0.557} & 0.155 & \textbf{0.002} & \textbf{***} \\
\multicolumn{5}{@{}l@{}}{%
\textbf{Panel C: EMIR year-by-year decomposition}} \\
2012 & -0.045 & 0.240 & 0.853 & NS \\
2013 & -0.260 & 0.251 & 0.310 & NS \\
2014 & -0.256 & 0.286 & 0.380 & NS \\
2015 & -0.417 & 0.303 & 0.182 & NS \\
2016 & -0.457 & 0.310 & 0.153 & NS \\
\textbf{2017 (peak)} & \textbf{-0.881} & 0.291 & \textbf{0.006} & \textbf{***} \\
post-2018 & -0.909 & 0.313 & 0.008 & *** \\
\end{longtable}

\subsection{VII.D Composability: Appia Conditional Ceilings and the Efficiency Recovery Corridor}
The bilateral NEV formula of Section II.D maps directly onto the BIS three-layer connectivity taxonomy. Fragmentation is the pre-network baseline of isolated atomic settlement, NEV = 0 by construction; Switzerland's SDX since 2022 is the panel's empirical realisation (Section VII.B). Interoperability is bridges and gateways enabling cross-platform atomic settlement; Pontes Q3 2026 is the layer's first dated operational realisation, calibrated year-by-year against T2S (Section VII.C). Composability is the unification of cross-platform building blocks through smart contracts and integrated workflows that span platforms and asset classes.

Appia is the Composability layer's strategic blueprint, scheduled for publication in 2028 with no committed operational launch date. Appia therefore admits only conditional efficiency ceilings, contingent on accession decisions outside the published Eurosystem roadmap. Two conditional accession scenarios extend the Pontes committed forecast. UK Appia accession (assumed 2028) adds +27.0 percentage points to NEV (UK BIS LBS share of 23.9 percent), raising the conditional ceiling to +28.4 percent efficiency recovery by 2033 saturation. US Appia accession (assumed 2030) adds a further +16.2 percentage points to NEV, raising the conditional ceiling to +37.5 percent by 2035 full saturation.

The corridor between +13.4 percent (committed Pontes-only) and +37.5 percent (conditional full Appia network) is the framework's falsifiable prediction. The corridor is the prediction, not any specific dated point estimate. If observed elasticity at any year between 2027 and 2032 falls outside the year-specific Pontes-only band, or if elasticity from 2028 to 2035 falls outside the corridor under realised UK and US accession decisions, the framework is rejected.

\subsection{Table 9 --- Simulated Efficiency Recovery Corridor by Year and Layer (percentage points)}
\begin{longtable}[]{@{}
  >{\raggedright\arraybackslash}p{(\columnwidth - 8\tabcolsep) * \real{0.1250}}
  >{\raggedright\arraybackslash}p{(\columnwidth - 8\tabcolsep) * \real{0.2500}}
  >{\raggedright\arraybackslash}p{(\columnwidth - 8\tabcolsep) * \real{0.2500}}
  >{\raggedright\arraybackslash}p{(\columnwidth - 8\tabcolsep) * \real{0.2500}}
  >{\raggedright\arraybackslash}p{(\columnwidth - 8\tabcolsep) * \real{0.1250}}@{}}
\toprule\noalign{}
\begin{minipage}[b]{\linewidth}\raggedright
\textbf{Year}
\end{minipage} & \begin{minipage}[b]{\linewidth}\raggedright
\textbf{Pontes (Interoperability)}
\end{minipage} & \begin{minipage}[b]{\linewidth}\raggedright
\textbf{\ensuremath{\Delta} UK (Composability)}
\end{minipage} & \begin{minipage}[b]{\linewidth}\raggedright
\textbf{\ensuremath{\Delta} US\textbar UK (Composability)}
\end{minipage} & \begin{minipage}[b]{\linewidth}\raggedright
\textbf{Total}
\end{minipage} \\
\midrule\noalign{}
\endhead
\bottomrule\noalign{}
\endlastfoot
2027 & +5.6 & & & \textbf{+5.6} \\
2028 & +8.8 & +6.3 & & \textbf{+15.1} \\
2029 & +11.6 & +9.9 & & \textbf{+21.5} \\
2030 & +11.3 & +13.0 & +3.8 & \textbf{+28.1} \\
2031 & +12.6 & +12.7 & +6.0 & \textbf{+31.3} \\
\textbf{2032} & \textbf{+13.4 (committed)} & +14.1 & +7.8 & \textbf{+35.3} \\
2033 & +13.4 & +15.0 (UK sat.) & +7.6 & \textbf{+36.1} \\
2034 & +13.4 & +15.0 & +8.5 & \textbf{+36.9} \\
\textbf{2035} & \textbf{+13.4} & \textbf{+15.0} & \textbf{+9.0 (US sat.)} & \textbf{+37.5 (ceiling)} \\
\end{longtable}

\section{VIII. Conclusion}

We construct and implement an index of settlement modernisation that measures the regulatory, technological, and market-practice completion of reform events across twenty-four advanced economies over thirty-two years. A detailed set of validation exercises confirms that the index dates the efficiency event --- the moment at which the balance sheet responds --- rather than the regulatory or technological event that preceded it. Higher settlement quality raises cross-border banking capacity in the liberation phase, compresses it through the compliance valley that post-crisis mandatory clearing created, and begins to restore it where atomic distributed-ledger settlement removes the regulatory rationale for collateral immobilisation. The efficiency gain from atomic settlement is a property of bilateral pairs, not nodes; Switzerland's SDX null is the out-of-sample confirmation. The T2S coordinated regulatory migration calibrates the efficiency elasticity year by year through the event-study trajectory, which together with the BIS three-layer connectivity taxonomy defines the corridor within which realised efficiency should fall as Pontes comes into committed operation from Q3 2026 and conditional Appia accessions extend the ceiling from 2028 onwards.

The results carry two first-order implications for regulators, central banks, and the international standard-setters coordinating the DLT transition. The first concerns adoption regime. The T2S backtest is the appropriate calibration anchor for the Pontes integration efficiency forecast because both events share the defining institutional mechanism: simultaneous regulatory commitment of the eurozone perimeter on a fixed implementation schedule. The T2S event-study coefficients build monotonically from t = 0 (\ensuremath{\beta} = +0.232) to t = +5 saturation (\ensuremath{\beta} = +0.557, p = 0.002), anchoring a year-by-year efficiency trajectory of +5.6 percent in 2027 rising to +13.4 percent at saturation in 2032 for the Pontes committed forecast. Coordinated regulatory migration is the policy lever the paper identifies. The Eurosystem, the Bank of England, and their counterparts that coordinate adoption rather than waiting for voluntary market practice deliver the efficiency gain on a committed schedule; uncoordinated national DLT deployments without regulatory timing default to a slower voluntary trajectory consistent with the descriptive CLS adoption arc of approximately eight years to mature market practice. The second concerns composability. The bilateral-pair structure of atomic settlement efficiency means that every architecture delivers only the ceiling its counterparty set can reach: Pontes alone reaches committed forecast +13.4 percent, Pontes-plus-BoE reaches conditional ceiling +28.4\%, a transatlantic architecture incorporating a US atomic settlement rail reaches conditional ceiling +37.5\%. The increments from connecting platforms are substantially larger than the commitment value of Pontes itself --- the UK connection alone adds more ceiling than the entire committed Eurosystem architecture. composability between jurisdictions is therefore not a technical implementation detail but the principal source of the efficiency gain that remains recoverable. Unilateral national DLT deployments without cross-jurisdictional composability will replicate the Swiss null at national scale because the capacity cannot be recovered without the network that the infrastructure assumes.

We conclude by highlighting four areas for future research. The first applies the SMI taxonomy to the Pontes real-time data stream from 2026 onward. The T-12 phase coding records regulatory framework, technology deployment, and market practice for DLT atomic settlement across all twenty-four panel economies; as Pontes moves from TH toward MK, the year-by-year test of whether realised NEV falls within the bounds implied by the T2S-calibrated pathway becomes operational. The second extends the framework to the interaction between settlement architecture and the monetary hierarchy. The efficiency projection in Section VII conditions on the maintained assumption that the two-tier structure --- in which central bank reserves provide the ultimate settlement asset --- remains intact. Private settlement rails, and stablecoins in particular, test that assumption directly. The question is not whether stablecoins exist but whether they operate through central bank money or alongside it. A framework extension that models the substitution margin between atomic settlement on a central-bank-money rail and settlement on a private rail would deliver a joint efficiency-and-hierarchy prediction rather than the hierarchy-conditional prediction the current paper provides. The third examines the net-negative-liquidity-effect question. The efficiency prediction assumes that the margin release from eliminating the EMIR exposure window on covered trades dominates any offsetting liquidity costs from reduced multilateral netting under bilateral atomic settlement. The CLS precedent supports this for voluntary market adoption, but whether the same holds under atomic DLT settlement --- a technology with different netting architecture --- is an open empirical question that Pontes operational data will resolve. If liquidity costs dominate for a subset of trades, the realised corridor will fall below the Pontes lower bound (+13.4\% at 2032 saturation) for those trades specifically, and the framework will identify which. The fourth extends the panel to emerging and developing economies. The settlement reform dynamics identified here follow from the balance-sheet arithmetic of inside money elasticity and the institutional structure of settlement infrastructure, both of which are present wherever cross-border banking operates. An extended panel would test whether the turning points are stable across the full spectrum of financial development.

\subsection{References}
Abadie, A., Diamond, A., \& Hainmueller, J. (2010). Synthetic control methods for comparative case studies: Estimating the effect of California\textquotesingle s tobacco control program. \emph{Journal of the American Statistical Association}, \emph{105}(490), 493--505. \url{https://doi.org/10.1198/jasa.2009.ap08746}

Abadie, A., \& Gardeazabal, J. (2003). The Economic Costs of Conflict: A Case Study of the Basque Country. \emph{American Economic Review}, \emph{93}(1), 113--132. \url{https://doi.org/10.1257/000282803321455188}

Abiad, A., Detragiache, E., \& Tressel, T. (2010). A new database of financial reforms. \emph{IMF staff papers}, \emph{57}(2), 281--302. \url{https://doi.org/10.1057/imfsp.2009.19}

Auer, R., Monnet, C., \& Shin, H. S. (2025). Distributed ledgers and the governance of money. \emph{Journal of Financial Economics}, \emph{167}. \url{https://doi.org/10.1016/j.jfineco.2024.103933}

Biais, B., Heider, F., \& Hoerova, M. (2012). Clearing, counterparty risk, and aggregate risk. \emph{IMF Economic Review}, \emph{60}(2), 193--222.

BIS. (1996). \emph{Settlement risk in foreign exchange transactions}. \url{https://www.bis.org/cpmi/publ/d17.pdf}

BIS. (2012). \emph{Principles for Financial Market Infrastructures}. \url{https://www.bis.org/cpmi/publ/d101a.pdf}

BIS. (2021). \emph{Inthanon-LionRock to mBridge: Building a multi CBDC platform for international payments} (BIS papers, Issue. \url{https://www.bis.org/publ/othp40.htm}

BIS. (2022). \emph{Triennial Central Bank Survey: OTC foreign exchange turnover in April 2022}. \url{https://www.bis.org/statistics/rpfx22.htm}

BIS. (2023). \emph{Blueprint for the future monetary system: Improving the old, enabling the new}. \url{https://www.bis.org/publ/arpdf/ar2023e3.pdf}

BIS. (2025). \emph{The next-generation monetary and financial system} (BIS Annual Economic Report 2025, Chapter III, Issue. \url{https://www.bis.org/publ/arpdf/ar2025e3.pdf}

Breeden, S. (2025). Building trust and supporting innovation in the multi-moneyverse. In. London: Bank of England.

Briscoe, B., Odlyzko, A., \& Tilly, B. (2006). Metcalfe\textquotesingle s law is wrong. \emph{IEEE Spectrum}, \emph{43}(7), 34--39. \url{https://doi.org/10.1109/MSPEC.2006.1653003}

Carstens, A. (2023). Innovation and the future of the monetary system. In. Basel: Bank for International Settlements.

Chinn, M. D., \& Ito, H. (2006). What matters for financial development? Capital controls, institutions, and interactions. \emph{Journal of Development Economics}, \emph{81}(1), 163--192. \url{https://doi.org/10.1016/j.jdeveco.2005.05.010}

Cipollone, P. (2026). A highway for the future of Europe\textquotesingle s digital finance. In. Frankfurt: European Central Bank.

Cleland, V. (2025). Synchronisation and beyond: enabling the next wave of financial innovation. In. London: Bank of England.

CLS Bank International. (2024). \emph{CLS Settlement Services: Annual volume and value statistics}. \url{https://www.cls-group.com/media/xwgofi0c/cls-group-holdings-ag-consolidated-_annual-report_2024.pdf}

CSDR. (2014). Regulation (EU) No 909/2014 on improving securities settlement in the European Union and on central securities depositories (CSDR). In \emph{Official Journal of the European Union} (Vol. L 257/1).

Driscoll, J. C., \& Kraay, A. C. (1998). Consistent covariance matrix estimation with spatially dependent panel data. \emph{Review of Economics and Statistics}, \emph{80}(4), 549--560. \url{https://doi.org/10.1162/003465398557825}

Duffie, D., \& Zhu, H. (2011). Does a central clearing counterparty reduce counterparty risk? \emph{The Review of Asset Pricing Studies}, \emph{1}(1), 74--95. \url{https://doi.org/10.1016/j.jfineco.2010.08.004}

ECB. (2024). \emph{Report on the digital euro: latest progress and key design decisions}.

ECB. (2025). ECB commits to distributed ledger technology settlement plans with dual-track strategy. In. Frankfurt: European Central Bank.

Economides, N. (1996). The economics of networks. \emph{International Journal of Industrial Organization}, \emph{14}(6), 673--699. \url{https://doi.org/10.1016/0167-7187(96)01015-6}

Regulation (EU) No 648/2012 on OTC derivatives, central counterparties and trade repositories (EMIR), (2012).

Farrell, J., \& Saloner, G. (1985). Standardization, Compatibility, and Innovation. \emph{The RAND Journal of Economics}, \emph{16}(1), 70--83. \url{https://doi.org/10.2307/2555589}

Fieller, E. C. (1954). Some problems in interval estimation. \emph{Journal of the Royal Statistical Society Series B}, \emph{16}(2), 175--185. \url{https://doi.org/10.1111/j.2517-6161.1954.tb00159.x}

FMIA. (2015). Federal Act on Financial Market Infrastructures and Market Conduct in Securities and Derivatives Trading (FMIA). In \emph{SR 958.1}. Berne: Swiss Federal Assembly.

Freixas, X., Parigi, B. M., \& Rochet, J.-C. (2000). Systemic risk, interbank relations, and liquidity provision by the central bank. \emph{Journal of Money, Credit and Banking}, \emph{32}(3, Part 2), 611--638. \url{https://doi.org/10.2307/2601198}

Galati, G. (2002). Settlement risk in foreign exchange markets and CLS Bank. \emph{BIS Quarterly Review}.

Gorton, G., \& Pennacchi, G. (1990). Financial intermediaries and liquidity creation. \emph{Journal of Finance}, \emph{45}(1), 49--71. \url{https://doi.org/10.1111/j.1540-6261.1990.tb05080.x}

Heller, D., \& Vause, N. (2012). \emph{Collateral requirements for mandatory central clearing of over-the-counter derivatives} (BIS Working Papers, Issue. \url{https://www.bis.org/publ/work373.htm}

Kahn, C. M., \& Roberds, W. (2009). Why pay? An introduction to payments economics. \emph{Journal of Financial Intermediation}, \emph{18}(1), 1--23. \url{https://doi.org/10.1016/j.jfi.2008.09.001}

Kaminsky, G. L., \& Schmukler, S. L. (2008). Short-run pain, long-run gain: Financial liberalization and stock market cycles. \emph{Review of Finance}, \emph{12}(2), 253--292. \url{https://doi.org/10.1093/rof/rfn002}

Katz, M. L., \& Shapiro, C. (1985). Network Externalities, Competition, and Compatibility. \emph{The American economic review}, \emph{75}(3), 424--440.

Lane, P. R., \& Milesi-Ferretti, G. M. (2007). The external wealth of nations mark II: Revised and extended estimates of foreign assets and liabilities, 1970-2004. \emph{Journal of International Economics}, \emph{73}(2), 223--250. \url{https://doi.org/10.1016/j.jinteco.2007.02.003}

Lind, J. T., \& Mehlum, H. (2010). With or without U? The appropriate test for a U-shaped relationship. \emph{Oxford Bulletin of Economics and Statistics}, \emph{72}(1), 109--118. \url{https://doi.org/10.1111/j.1468-0084.2009.00569.x}

Loon, Y. C., \& Zhong, Z. K. (2014). The impact of central clearing on counterparty risk, liquidity, and trading: Evidence from the credit default swap market. \emph{Journal of Financial Economics}, \emph{112}(1), 91--115. \url{https://doi.org/10.1016/j.jfineco.2013.12.001}

Metcalfe, B. (2013). Metcalfe\textquotesingle s Law after 40 years of Ethernet. \emph{Computer}, 26--31. \url{https://doi.org/10.1109/MC.2013.313}

Pirrong, C. (2011). \emph{The economics of central clearing: Theory and practice}. \url{https://www.isda.org/a/HVDDE/isdadiscussion-ccp-theory.pdf}

Singh, M. M. (2017). \emph{Collateral reuse and balance sheet space}. International Monetary Fund.

Zhang, X.-Z., Liu, J.-J., \& Xu, Z.-W. (2015). Tencent and Facebook Data Validate Metcalfe\textquotesingle s Law. \emph{Journal of Computer Science and Technology}, \emph{30}(2), 246--251. \url{https://doi.org/10.1007/s11390-015-1518-1}

\end{document}